\documentclass[twocolumn]{aastex631}

\shorttitle{Pearls on a String}
\shortauthors{Coronado et al.}
\graphicspath{{./}{figs/}}

\begin{document}

\title{Pearls on a String: Numerous Stellar Clusters Strung along the Same Orbit}

\correspondingauthor{Johanna Coronado}
\email{coronado@mpia.de, jca.coronadomartinez@gmail.com}

\author[0000-0001-6629-0731]{Johanna Coronado}
\affiliation{Max-Planck-Institut f\"ur Astronomie \\
K\"onigstuhl 17, D-69117\\
Heidelberg, Germany}

\author[0000-0002-9460-0212]{Verena F\"urnkranz}
\affiliation{Max-Planck-Institut f\"ur Astronomie \\
K\"onigstuhl 17, D-69117\\
Heidelberg, Germany}

\author[0000-0003-4996-9069]{Hans-Walter Rix}
\affiliation{Max-Planck-Institut f\"ur Astronomie \\
K\"onigstuhl 17, D-69117\\
Heidelberg, Germany}



\begin{abstract}
Stars originate from the dense interstellar medium, which exhibits filamentary structure to scales of $\sim 1$~kpc in galaxies like our Milky Way. 
We explore quantitatively how much resulting large-scale correlation there is among different stellar clusters and associations in \emph{orbit phase space}, characterized here by actions and angles.  As a starting point, we identified 55 prominent stellar overdensities in the 6D space of orbit (actions) and orbital phase (angles), among the $\sim$ 2.8 million stars with radial velocities from Gaia EDR3 and $d \leq 800$~pc. We then explored the orbital \emph{phase} distribution of all sample stars in the same \emph{orbit} patch as any one of these 55 overdensities.  We find that very commonly numerous other distinct orbital phase overdensities exist along these same orbits, like pearls on a string. These `pearls' range from known stellar clusters to loose, unrecognized associations. Among orbit patches defined by one initial orbit-phase overdensity 50\% contain at least 8 additional orbital-phase pearls of 10 cataloged members; 20\% of them contain 20 additional pearls. This is in contrast to matching orbit patches sampled from a smooth mock catalog, or offset nearby orbit patches, where there are only 2 (or 5, respectively) comparable pearls. Our findings quantify for the first time how common it is for star clusters and associations to form at distinct orbital phases of nearly the same orbit. This may eventually offer a new way to probe the 6D orbit structure of the filamentary interstellar medium. 

\end{abstract}



\section{Introduction} \label{sec:intro}

Clusters form in hierarchically structured, accreting giant molecular clouds (GMCs), which are part of the overall filamentary structure of the cold interstellar medium.  Within these GMCs there are density concentrations known as clumps  \citep{2019ARA&A..57..227K}, where individual small regions  collapse independently (fragment) and eventually produce stars until the gas is dispersed by feedback. 

The densest parts of the hierarchy resist gas removal (until they reach a high star-formation efficiency) and they become dynamically relaxed and well mixed, where they remain bound after gas removal. In the first $\sim$100 Myr after the latter, clusters disperse moderately fast \citep{2019ARA&A..57..227K}. In this scenario, young star clusters are expected to preserve some memory of the physical conditions that are present in their birthplace \citep{2010MNRAS.403..996B}. 

Star clusters in the disk of the Milky Way are commonly classified as open clusters (OCs) or associations. And it has not yet been well quantified or understood to which extent distinct OCs and associations are born isolated or in correlated ensembles. Although, gravitating matter tends to cluster and form objects on different scales \citep{2020A&A...642L...4K}. Thus it could be possible that, similar to stars, clusters may form groups and physical pairs \citep{1996ApJ...466..802E,1997AJ....113..249F,2010MNRAS.403..996B,2013MNRAS.434..313G}. 

In nearby external galaxies, such as the Large Magellanic Cloud (LMC) and Small Magellanic Cloud (SMC), where the viewing geometry is simpler, binary OCs or groups of OCs are observed. Some studies have shown that $\sim$ 12 per cent of clusters are in pairs or binary clusters \citep{2000AcA....50..355P}. In our own Galaxy, the fraction of binary clusters has been found to be approximately the same \citep{2009A&A...500L..13D}.

Binary clusters appear to be systematically younger than single clusters \citep{2002A&A...391..547D,2016A&A...586A..41P}, where nearly half of them are $<$ 25 Myr old \citep{2009A&A...500L..13D}. And clusters that form a binary are often coeval \citep{1993A&A...267...59K,2012ApJ...746L..19M}. Such cluster multiplicity could be a natural consequence of the cold ISM structure \citep{1996ApJ...471..816E}.

It is been known for $\sim$ 50 years that some widely separated clusters lie on essentially the same orbit. For example, the Hyades and Praesepe have similar kinematic properties \citep{1959Obs....79..143E,1960MNRAS.120..540E}, and are part of the Hyades supercluster \citep{2015ApJ...807...24B}. It has also long been thought that these clusters are coeval, having formed from a single molecular cloud, or cloud complex \citep{2013ApJ...775...58B}. Both clusters seem to have similar (literature) ages, as shown by isochrone fitting \citep{1998A&A...331...81P,2004A&A...414..163S} and gyrochronology \citep{2014ApJ...795..161D,2014MNRAS.442.2081K}. Moreover, their metallicities are very similar at around [Fe/H] $\sim$ +0.13 \citep{2013ApJ...775...58B}. 

Gaia data now allow us to address more systematically what the correlation or connection among distinct clusters and associations is in \emph{orbit space}.  
This could help us to understand the both formation of stars, the large-scale dynamics of GMC cores, and the structure and dynamics of our Galaxy.  \citet{2017A&A...600A.106C} used a sample of known open clusters with 6D (pre-Gaia) phase-space information to group them via a $(\vec{x},\vec{v})$-space friends-of-friends algorithm (FoF).  They found 14 clusters pairs and a handful of larger groups of clusters. And they showed that those were more FoF groupings than expected at random. Here, we build on this initial result by carrying out a more far-reaching analysis of orbit correlations among star clusters and associations, using action-angle space coordinates and drawing on Gaia EDR3 data \citep{2020arXiv201203380L}.

When going beyond the immediate neighbourhood (say, 50~pc)  action--angles ($J,\theta$) coordinates stand out as the canonical choice to describe orbits\footnote{Orbits may be fully described by actions and angles in axisymmetric potentials. In the presence of non-axisymmetric perturbations, such as spiral arms, actions can still serve as approximate orbit labels \citep{Trick_2017}. For volumes around the solar neighbourhood that are larger than $\sim~100$ pc, actions are preferable over commonly used alternatives such as space motions UVW, which are already only approximate orbit labels in axisymmetric potentials.}; this is because orbital motions take on a particularly simple function form: actions are conserved, and angles (or, orbital phases) grow linearly with time \citep{binney2008}. Moreover, actions are adiabatic invariants under slow changes of the gravitational potential, gradual orbit evolution may be described as a diffusion in action space \citep{2015MNRAS.449.3479S}. Actions and angles are also a powerful coordinate system to find more general substructure in the orbit distribution of our Galaxy, where they have been used to identify systematically groups of stars on similar orbits (e.g., \citet{2019MNRAS.484.3291T,2020MNRAS.495.4098C}), even if they are spatially dispersed. 

In this paper, we now set out to address the following question: on orbits in the Galactic disk, selected to contain at least one cluster or association, how much more likely is it to find other clusters and associations, compared to offset adjacent orbits? 

Conceptually, we do this as follows: We take the sample of Gaia EDR3 stars within 800~pc of the Sun that have 6D phase-space and hence action-angle coordinates, those that have radial velocities. From those we identify 55 prominent overdensities, distinct groups of stars whose members are exceptionally close to one another in orbit \emph{and} orbital phase, following closely the action-angle approach of \citet{2020MNRAS.495.4098C}. Among them are of course a number of known clusters (e.g. M67, Praesepe and the Pleiades, but also stream-like groups that are very extended across the sky, such as the Pisces Eridanus or Meingast 1 stream \citep{2019A&A...622L..13M}). For each of these 55 orbit -- phase overdensities, we then consider \emph{all} sample stars in the same orbit patch (i.e. at nearly the same action), irrespective of their phase along the orbit. We then look at their distribution in orbital phase.  By construction there must be at least one orbit-phase overdensity, namely that which defined the orbit patch. However, our results reveal that commonly there are multiple other orbit-phase overdensities (clusters or associations) on the same orbit. They line up at different orbit phase along the same orbit, like ``pearls on a string''. To put this into context, we perform an analogous analysis for a mock-catalog without clusters \citep{2020PASP..132g4501R} and for an orbit patch that is offset from the actions where the initial action-angle density was detected. The goal is then to quantify statistically any differences between the incidence of such  orbit-phase overdensities (pearls) on orbits that are centered on one known pearl, compared to orbits that are not. Those differences turn out to be quite dramatic.

The paper is organised as follows. Section~\ref{sec:data} presents the data used in this study, from Gaia EDR3 and the Gaia EDR3 mock catalog. Section~\ref{sec:clumps_actions_angles} reviews our method for finding groups in action-angle space, and then describes how we quantify overdensities in orbital-phase space at a given orbit patch. In Section~\ref{sec:results_disc} we first show three case studies in some detail, and then present the statistics of pearls on orbits known to contain at least one pearl to offset similar disk-like orbits. Finally, we summarize the results in Section~\ref{sec:summary}. Additional information and plots can be found in the Appendix.

\section{Data}
\label{sec:data}
\subsection{The Gaia EDR3-RVS Sample}

For the analysis of this paper we make use of the full 6D information ($\alpha, \delta, \varpi, \mu_{\alpha\star}, \mu_{\delta}, v_{r}$) from the Gaia Early third Data Release (EDR3, \citet{2020arXiv201203380L}), with $\mu_{\alpha\star}=\mu_{\alpha}cos\delta$. We denote the EDR3 subsample with useable RVS radial velocity measurements as EDR3-RVS. 
Compared to Gaia DR2, EDR3 affords a $\sim$ 30\% and $\sim$ 100\% improvement in parallax precision and proper motions, respectively. For this paper, we use of the photogeometric distances derived by \citet{2020arXiv201205220B}. We impose the following quality cuts to this dataset: \texttt{ruwe} $< 1.4$ and formally $\varpi/\sigma_\varpi > 3$ \citep{2020arXiv201206242F}; in practice, $> 99.99\%$ of all stars have $\varpi/\sigma_\varpi > 20$, making $\varpi/\sigma_\varpi > 20$ the effective sample cut.
We restrict ourselves to stars within $d_{photo-geo} \leq 800$ pc, leaving 2.4 million stars. This distance limit still leaves us with a volume that should contain much sub-structure, while limiting the impact of imprecise distances when calculating actions \citep{2018MNRAS.481.2970C}. 

In using the full EDR3-RVS we must forego the use of chemical abundances in stellar clustering analyses, as in \citet{2020MNRAS.495.4098C}. We do this, because \textit{a)} this sample offers an independent confirmation of the action-angle groups found in \citet{2020MNRAS.495.4098C}; \textit{b)} it is a vastly larger sample ($\times$10) than the latter and \textit{c)} with this sample we will also have an all-sky coverage, as we are not limited to a spectroscopic survey. The absence of stellar abundances matters little, as young groups ($<1-2$~Gyrs) in the solar neighborhood have almost exclusively metallicities similar to solar metallicity \citep{2017A&A...601A..70S}.

\subsection{A Mock Catalog with a Smooth and Phase-Mixed Orbit Distribution}
\label{sec:gaia_mock}

As a null hypothesis for our analysis clustering in orbital phase, we want an analogous data set, but with smooth and fully phase-mixed orbit distribution. We do this by creating a `mock sample' that matches our Gaia EDR3 selection in volume and depth, based on the Gaia early DR3 mock stellar catalog by \citet{2020PASP..132g4501R}. We query stars with 
 4 $<$ \texttt{phot\char`_mean\char`_mag} $<$ 13 and 3500 $<$\texttt{teff} $<$ 6900, as this is the temperature and photometry range for stars in the radial velocity sample \citep{2019A&A...622A.205K}. We invert the parallaxes that gives us exact model distances in this mock catalog \citep{2018PASP..130g4101R} and we select stars within $d \leq 800$ pc. To add realism, the mock catalog \citet{2020PASP..132g4501R} includes moving groups and clusters, which we remove for his application by including \texttt{WHERE popid! = 11} in the query. We calculate actions and angles from the exact values without adding uncertainties. 

\section{Quantifying clumping in action-angle space: Pearls}
\label{sec:clumps_actions_angles}

For the analysis approach sketched in the introduction we need to quantify clumping in both 6D action-angle space, and in 3D angle space, to identify compact stellar overdensities. First in orbit and phase space, then in orbital phase for all stars at a given orbit space patch. To do so, we follow closely the method described in Sec. 3 from \citet{2020MNRAS.495.4098C}, based on the friends-of-friends approach. 

\subsection{Action-angle computation}
\label{sec:action_angle_comp}
The calculation of actions {\bf J} and angles $\theta$ require both ($\mathnormal{x,v}$) phase-space coordinates, and a gravitational potential. By assuming that the Galaxy's potential is close to an axisymmetric St{\"a}ckel potential, one can easily calculate actions and angles for the stars in our sample. We make use of the python package \texttt{galpy}, with its implementation of the action estimation algorithm \emph{St{\"a}ckel fudge} \citep{2012MNRAS.426.1324B} along with the \texttt{MWPotential2014} model. The latter considers a simple axisymmetric Milky Way potential model with a circular velocity of 220 km/s at the solar radius of 8 kpc \citep{galpy}. Note that the absolute values of the actions never enter the subsequent analysis, just their differences. So, the choice of an updated circular velocity (e.g., \citet{2019ApJ...871..120E}) would not significantly alter the results.

 For the location and velocity of the Sun within the Galaxy we assume $(X,Y,Z)$ = (8,0,0.025) kpc and $(U,V,W)_{\odot}$ = (11.1,12.24,7.25) km/s \citep{2010MNRAS.403.1829S} to first calculate Galactocentric coordinates and then actions from the observed ($\alpha, \delta, d, v_{r}, \mu_{\alpha\star},\mu_{\delta}$) of each star. For our sample, which has been restricted to $d \le 800$ pc and $\varpi/\sigma_\varpi > 20$, the largest contribution to the action uncertainties arise from the radial velocity uncertainties.

\subsection{Finding action-angle groups in GEDR3}
\label{sec:fof_algorithm}

 Any friends-of-friends approach requires a scalar distance measure between any two stars, which is then compared to the linking length to see whether they are `friends'; this in turn requires a 6D or 3D distance metric. To define such a metric, we follow \citet{2020MNRAS.495.4098C} where it is defined via a diagonal metric tensor whose elements are determined by the inverse variance of stars' positions in each of the coordinates (see Section 3.3. and  Eqs. 1--5 from \citet{2020MNRAS.495.4098C}). This yields a linking length, $l=\log_{10} \Delta(J,\theta)$ in action-angle space, expressed as the (logarithm of the) fraction of the median 6D separation, $\Delta(J,\theta)$ between sample stars.
For any choice of $l$, we consecutively join all distinct pairs closer than $l$ that have a star in common. This results in a number of associations of $\ge 3$ members for any linking length; remaining isolated pairs are discarded from further consideration. This FoF approach is strictly algorithmic for a given $l$, and in \citet{2020MNRAS.495.4098C} we showed that we can identify known clusters, as well as known and new associations, even if they are spread across the entire sky. However, one must choose a linking length $l$. We explored a range of values for it ($log_{10}\,\Delta(J,\theta$):  $l_i = [-1.8, -1.7, -1.6]$),
and adopted $log_{10}\,\Delta(J,\theta) = -1.7$; for more details on the role of the linking length in choosing such groups, see also \citep{2020MNRAS.495.4098C}.

For the analysis presented in this paper we selected the richest groups identified for this $l$, as defined by their minimal number of members required: 55 groups with $\geq20$ members each\footnote{Full Table~\ref{tab:vizier} is available in a machine-readable format in the online Journal.}. These are prominent overdensities in action-angle space, stars on nearly the same orbit at nearly the same orbital phase. 
It should be noted that the number of these FoF groups here (55) is of course much smaller than the number of known clusters or associations in the extended Solar neighbourhood. This is because we require at least 20 members with Gaia RVS velocities. The bright magnitude limit of RVS in EDR3 and the incompleteness for RVS velocities in crowded fields cause many well-known clusters not to be included. This incompleteness of the EDR3-RVS catalog does not affect the subsequent analysis.

\begin{figure*}
\setlength{\unitlength}{\textwidth}
\begin{center}
\begin{picture}(1,1)
\put(0,0){\includegraphics[width=\textwidth]{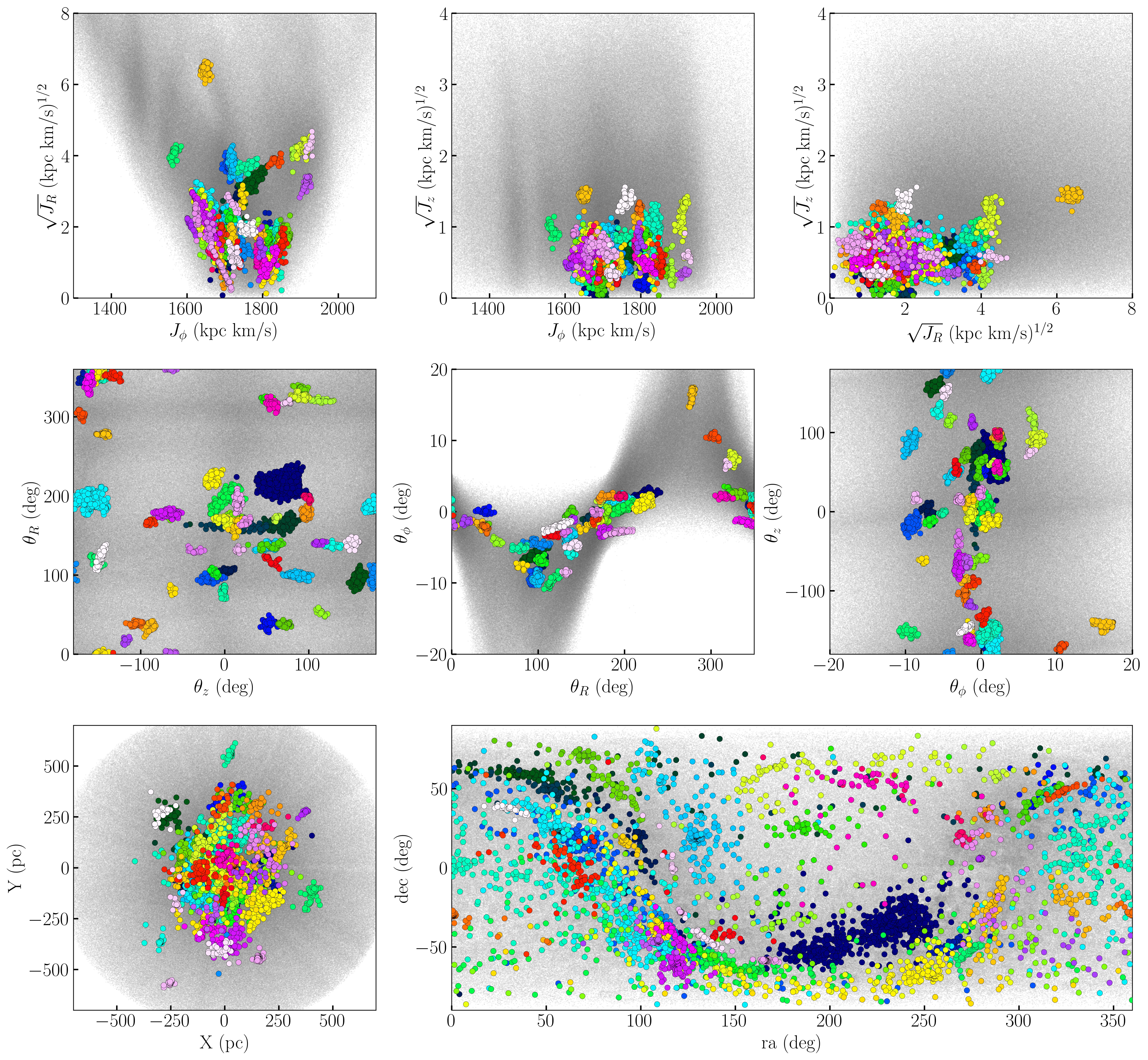}}
\end{picture}
\caption{The 55 groups in action-angle space with at least 20 members, identified by the FoF algorithm applied to the Gaia 6D data. The different panels show all the members of the (color-coded) 55 groups in different coordinates.
\newline
\textit{\textbf{Top and middle row}}: The distribution of the groups in the three projections of action and of angle space; by construction, all groups are tightly clustered in these coordinates.
\newline
\textit{\textbf{Bottom row}}:  Top-down view of these 55 groups in heliocentric Galactic (X,Y) coordinates (left), and the groups' on-sky distribution right ascension and declination (right), showing how extended many of them appear in these coordinates.}
\label{fig:all_groups}
\end{center}
\end{figure*}

Their basic properties ($N_{members}$ along with their mean actions and angles) are listed in Table~\ref{table:table1}, and we illustrate their distribution in Figure~\ref{fig:all_groups}: the top and middle row display the groups in action and angle space. By construction, each group forms a compact clump in both orbit and orbital phase space. The bottom row in Fig.~\ref{fig:all_groups} illustrates a top-down view of the groups in heliocentric Galactic coordinates (left), and the spatial distribution of the groups in right ascension and declination (right). Albeit the groups are compact in action angle space, they are spatially more dispersed and extend over several 10s to 100s degrees on the sky.

We will now turn to the question whether in the same orbit patch (at the nearly the same action) there are other distinct overdensities in orbital phase.

\subsection{Defining orbit patches around an action-angle overdensity}
\label{sec:action_ellipsoid}
For any action-angle group of stars we consider, we want to identify all the stars in the entire sample that are on similar orbits {\bf J}, irrespective of their phases $\theta$. To this end, we first need to define an \textquotedblleft orbit patch \textquotedblright in 3D action space around the mean {\bf J} of the group we had originally found.

We do this by fitting a 3D ellipsoid to the distribution of actions ($\mathbf{J}~=~J_{R}, J_{z}, J_{\phi}$) of each identified group. These ellipsoids are defined by the centroid and covariance matrix of the members' {\bf J}, and describe the extent and orientation of the orbit patch in action space. To include all members, we chose to enlarge this ellipsoid to 3$\sigma$ when defining the orbit patch. This extent also allows the inclusion of most group members if the distribution in action space is not exactly Gaussian.

Throughout the next Sections, we illustrate the various analysis steps using three FoF groups, namely groups 12, 19 and 40 in Table~\ref{tab:long}. These three groups were chosen because they turned out to be typical for the entire sample with respect to the subsequent analysis.

\begin{figure*}
\setlength{\unitlength}{\textwidth}
\begin{center}
\begin{picture}(1,1)
\put(0,0.65){\includegraphics[width=\textwidth]{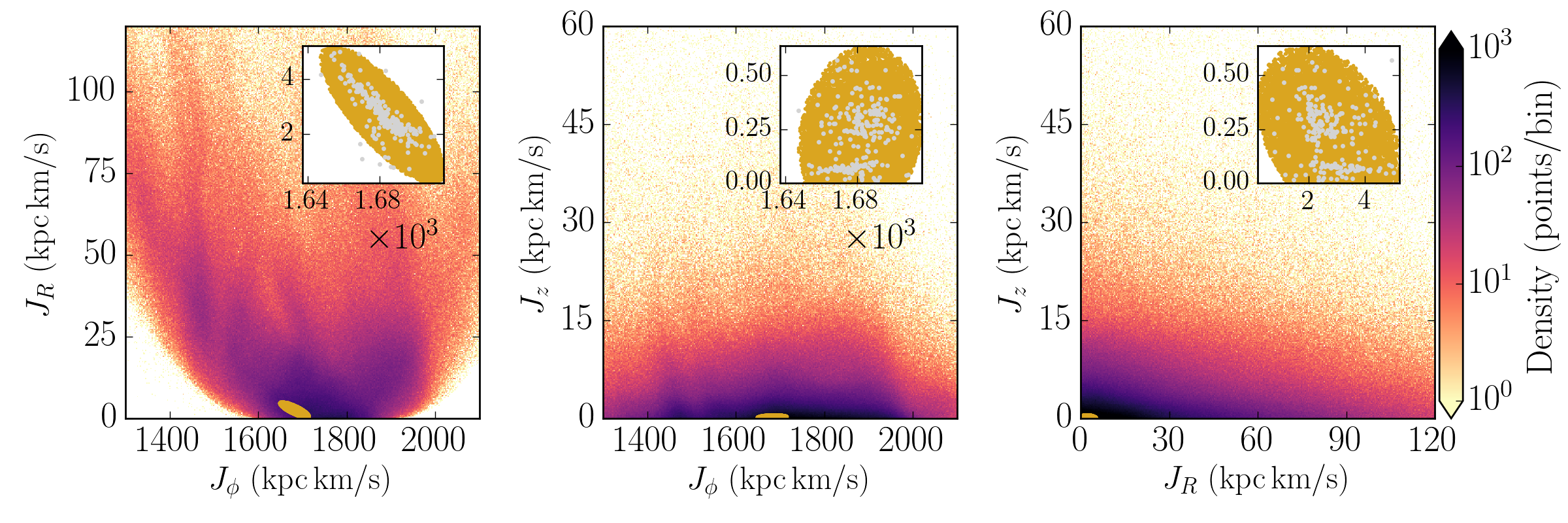}}
\put(-0.02,0){\includegraphics[width=\textwidth]{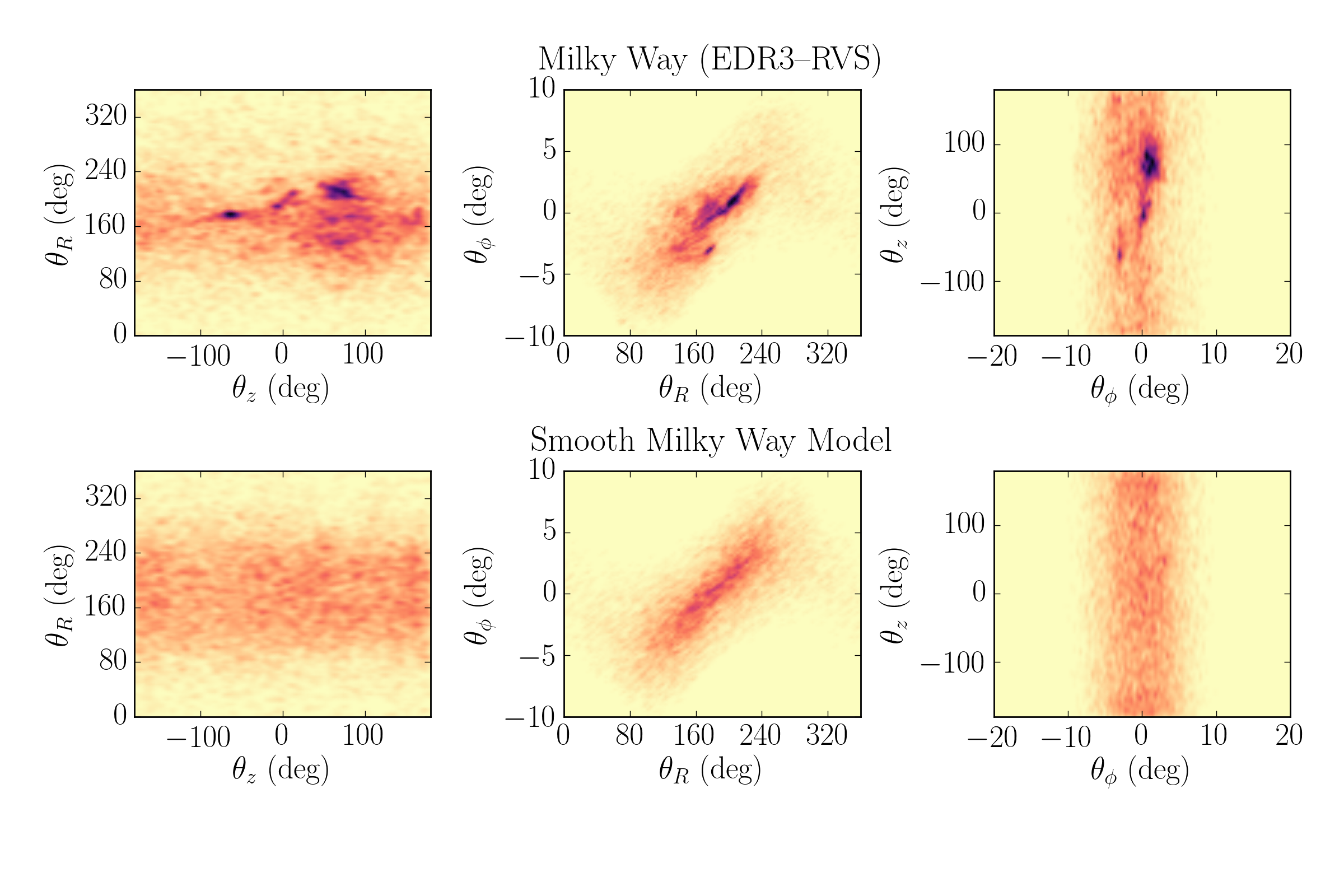}}
\end{picture}
\caption{Manifestations of `pearls on a string': the case of the \textit{FoF G19} group. The top row shows three projections of the orbital action distribution of the entire Gaia EDR3-RVS, $p({\bf J})~=~p(J_{R}, J_{z}, J_{\phi})$, with the orbit patch of one prominent action-angle overdensity , \textit{FoF G1}, highlighted, which had been defined via a 6D friends-of-friend (FoF) algorithm in orbital phase-space. The actions of the FoF-identified stars are shown as grey points in the inset; the orbit patch as a golden ellipse. In the middle row the color represents the 2D projection of the 3D kernel density estimate of the star-by-star distribution in orbital phase (or angle) $p(\theta_{R}, \theta_{z}, \theta_{\phi})$ for all stars in the \textit{FoF G19} orbit patch, ${\bf J}\approx {\bf J}_{G19}$.} The bottom row shows the analogous distribution in the same orbit patch, but with points drawn from a smooth mock catalog.  Both angle-distributions contain $\sim$12\ 700 stars, and are based on the same kernel bandwidth. The angle distribution for the observed stars in this \textit{FoF G19} orbit patch is highly structured, with the presence of multiple, distinct orbital-phase overdensities. These are distinct orbit-phase overdensities strung along the nearly the same orbit, like ``pearls on a string''. The angles defining the initial \textit{FoF G19} are located at ($\theta_{R}, \theta_{z}, \theta_{\phi}$) = (200, 0, 0) deg. In stark contrast, the angle distribution drawn from the mock catalog is no-uniform (reflecting spatial selection effects), but smooth.
\label{fig:ellipsoid_example}
\end{center}
\end{figure*}

Fig.~\ref{fig:ellipsoid_example} illustrates this approach for one example FoF group with 231 member stars (hereafter \textit{FoF G19}). After having determined {\bf J}, the covariance matrix and $\sigma$ we select every star from the Gaia EDR3-RVS whose actions {\bf J} lie within this patch, as shown by the yellow dots in Fig.~\ref{fig:ellipsoid_example}. For \textit{FoF G19} this selection comprises $\sim$ 12\ 700 stars, whose distribution of {\bf J} indeed reflect the {\bf J} of the original action-\emph{angle} group that we show with grey dots. 

\subsection{The angle distribution of stars within an orbit patch}
\label{sec:angles-in-FoF-patch}
We now explore the angle (orbital phase) distribution of all stars within an orbit patch. We start by looking at orbit patches centered on various FoF groups, then compare them to `offset' orbit patches.

For the orbit patch centered on \textit{FoF G19} these stars are the yellow dots in the top panel of Fig.~\ref{fig:ellipsoid_example}. The three projections of their 3D angle density distribution are illustrated in the lower panels of that same figure, represented by a Gaussian kernel density estimate. By construction, we expect to see at least one clump in angle space, the angles distribution of \textit{FoF G19} with its 231 members, which defined the orbit patch. However, the angle distribution unveils quite a number of distinct compact orbital phase overdensities in the survey volume of 800~pc around us.
These overdensities may include star clusters, associations or disk streams, so we need an encompassing terminology for them. And since they are are strung along the same orbit at different angles, like pearls on a string, we use the term 'pearls' for all these multiple angle overdensities in the same orbital patch. This generalizes the known notion that several clusters or associations can be found along (nearly) the same orbit.  

The lower panels of Fig.~\ref{fig:ellipsoid_example} show the angle distribution of the same orbit patch, but for data drawn from the smooth and phase-mixed Gaia mock catalog described in Section~\ref{sec:gaia_mock}. It reproduces the general features of the real angle distribution: the angle distribution is not uniform, as the finite sample volume -- the 800~pc around us -- selects particular angles for a given orbit. But the angle distribution drawn from the mock catalog shows no localised clustering, as expected for a smooth phased-mixed distribution. Hence, for a smooth orbital phase distribution (except \textit{FoF G19}) in the top panel we would have expected to see just one distinct orbital-phase overdensity. 

\begin{figure*}
\begin{center}
\includegraphics[width=\textwidth]{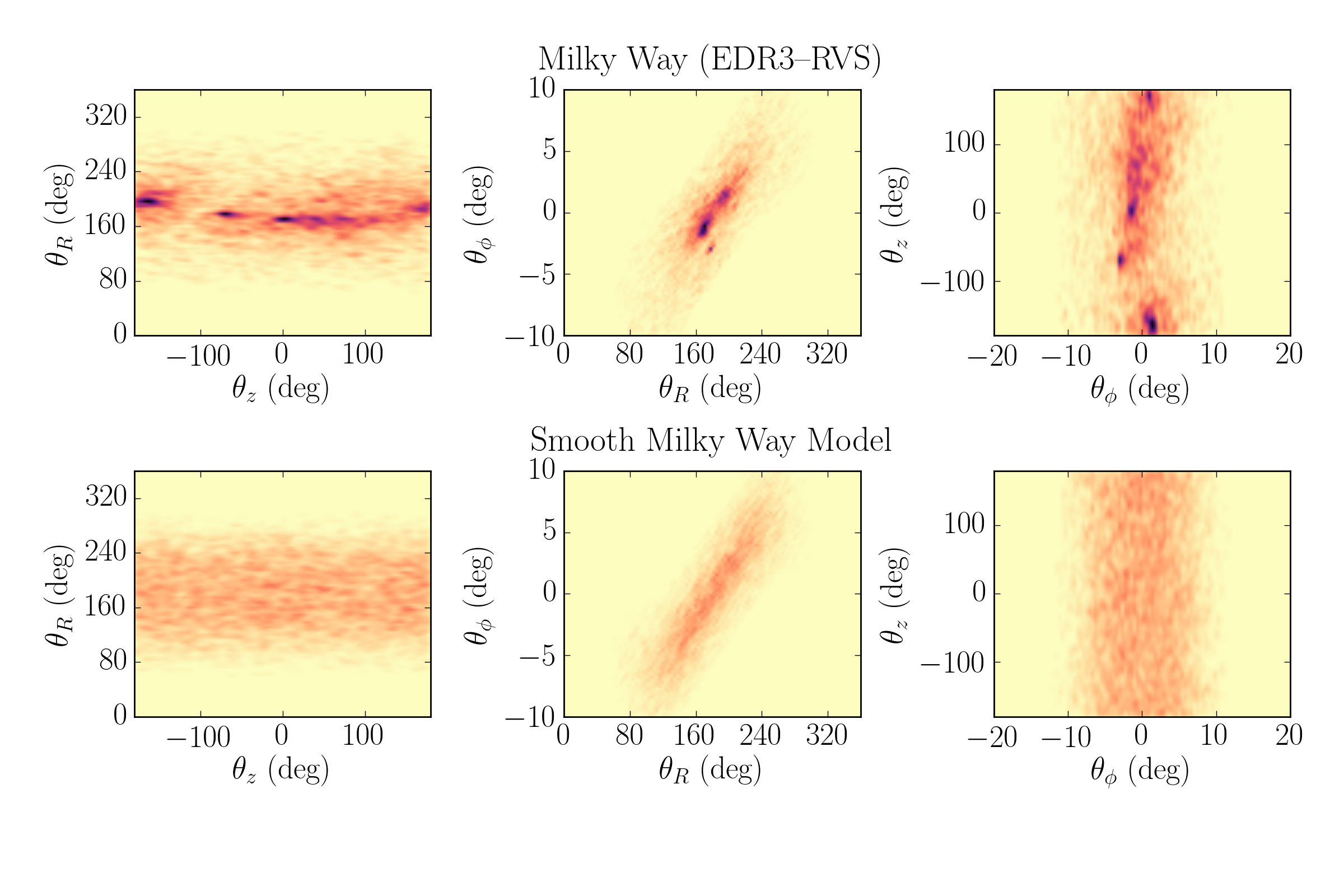}
\caption{Manifestations of `pearls on a string': the case of the \textit{FoF G40} group. This figure panel is analogous to the two bottom rows of Figure~\ref{fig:ellipsoid_example}, but here for the \textit{FoF G40} overdensity as a second case study. The top row here shows the projections of the orbital phase (or angle) distribution $p(\theta_{R}, \theta_{z}, \theta_{\phi})$ (using a kernel density map)  for all stars in the \textit{FoF G40} orbit patch, ${\bf J}\approx {\bf J}_{G40}$. The bottom row shows the analogous distribution in the same orbit patch, but with points drawn from a smooth mock catalog.  Both angle-distributions contain $\sim$5700 stars, and are based on the same kernel bandwidth. The angle distribution for the observed stars in this \textit{FoF G40} orbit patch is highly structured, with the presence of multiple, distinct orbital-phase overdensities, which we dub `pearls'. The angles defining the initial \textit{FoF G40} is approximately located at ($\theta_{R}, \theta_{z}, \theta_{\phi}$) = (170, 5, -1) deg, its exact position is shown in the middle panel of Fig.~\ref{fig:angles_G1_G2_G3}. Again in contrast, the angle distribution drawn from the mock catalog is no-uniform (reflecting spatial selection effects), but smooth.}
\label{fig:angle_ellipsoid_example2}
\end{center}
\end{figure*}

\begin{figure*}
\begin{center}
\includegraphics[width=\textwidth]{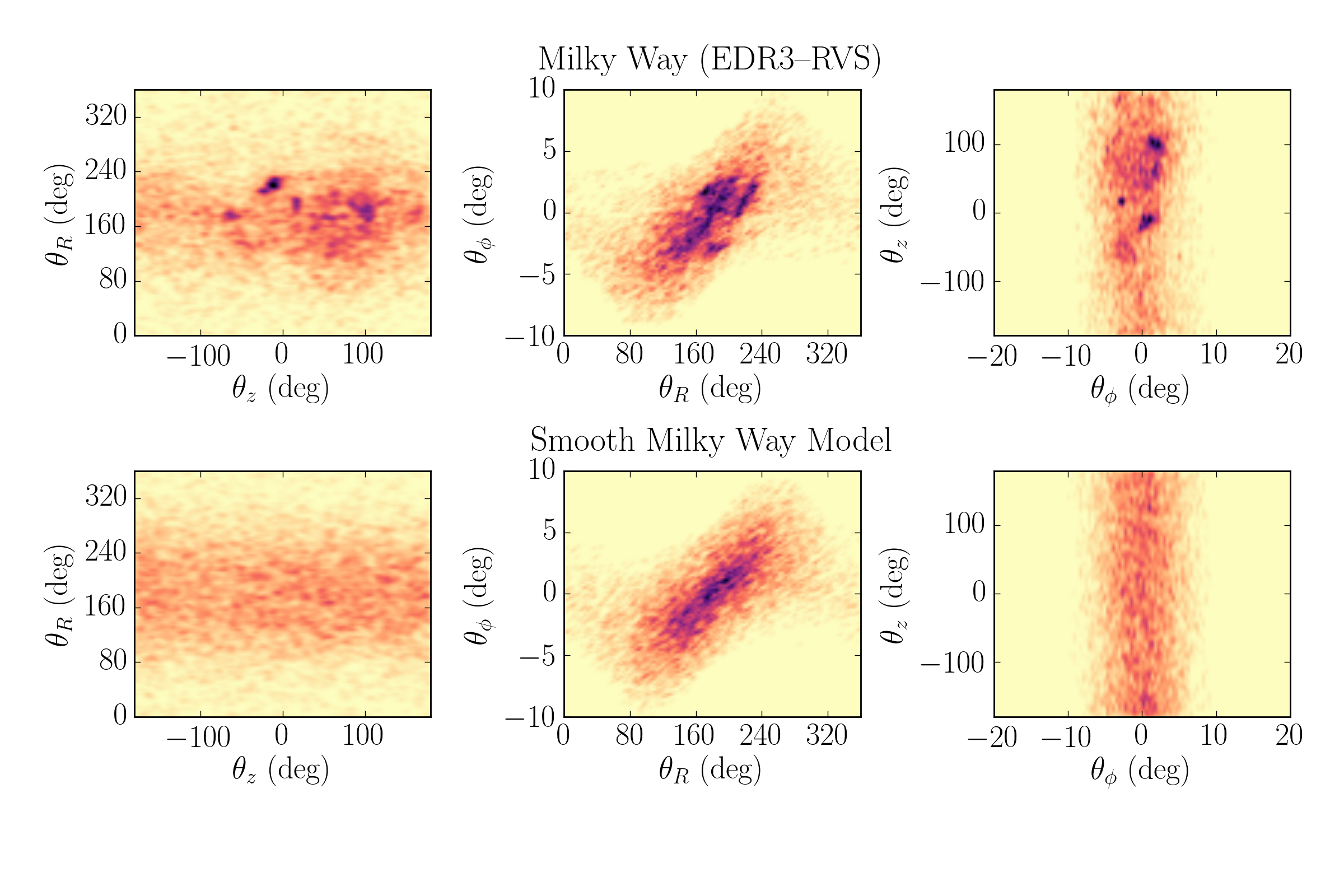}
\caption{Manifestations of `pearls on a string': the case of the \textit{FoF G12} group. This figure panel is analogous to Figure~\ref{fig:angle_ellipsoid_example2}, but here for the \textit{FoF G12} overdensity as a final case study. The top row here shows a the projected kernel density map of the orbital phase (or angle) distribution $p(\theta_{R}, \theta_{z}, \theta_{\phi})$ for all stars in the \textit{FoF G12} orbit patch, ${\bf J}\approx {\bf J}_{G12}$. The bottom row shows the analogous distribution in the same orbit patch, but with points drawn from a smooth mock catalog.  Both angle-distributions contain $\sim$7900 stars, and are based on the same kernel bandwidth. The angle distribution for the observed stars in this \textit{FoF G12} orbit patch is highly structured, with the presence of multiple, distinct orbital-phase overdensities, which we dub `pearls'. The angles defining the initial \textit{FoF G12} is approximately located at ($\theta_{R}, \theta_{z}, \theta_{\phi}$) = (230, -5, 1) deg and its exact position is shown in Fig.~\ref{fig:angles_G1_G2_G3}. Again in contrast, the angle distribution drawn from the mock catalog is no-uniform (reflecting spatial selection effects), but smooth.}
\label{fig:angle_ellipsoid_example3}
\end{center}
\end{figure*}

We then examine the analogous orbital phase distributions for the orbit patches centered on the remaining 54 FoF groups. Two more examples are shown Figures~\ref{fig:angle_ellipsoid_example2} and~\ref{fig:angle_ellipsoid_example3} (hereafter \textit{FoF G40} and \textit{FoF G12}, respectively), along with their corresponding orbit patch selections in Figs.~\ref{fig:A_ellipsoid_example2} and~\ref{fig:A_ellipsoid_example3} in the Appendix. These two figures show once again that when selecting all stars in an orbit patch selected for the existence of one action-angle group, the distribution in angle space is highly structured showing: we see many more orbital-phase overdensities than just the one belonging to the selected group. Again, the lower panel in both figures shows the comparison for the same orbit patch to a smooth phased-mixed distribution, highlighting the general features of the data when there is no clustering.

\subsubsection{The orbital phase distribution in \textquoteleft offset\textquoteright\,orbit patches}
\label{sec:random_ellipsoid}

\begin{figure*}
\setlength{\unitlength}{\textwidth}
\begin{center}
\begin{picture}(1,1)
\put(0,0.65){\includegraphics[width=\textwidth]{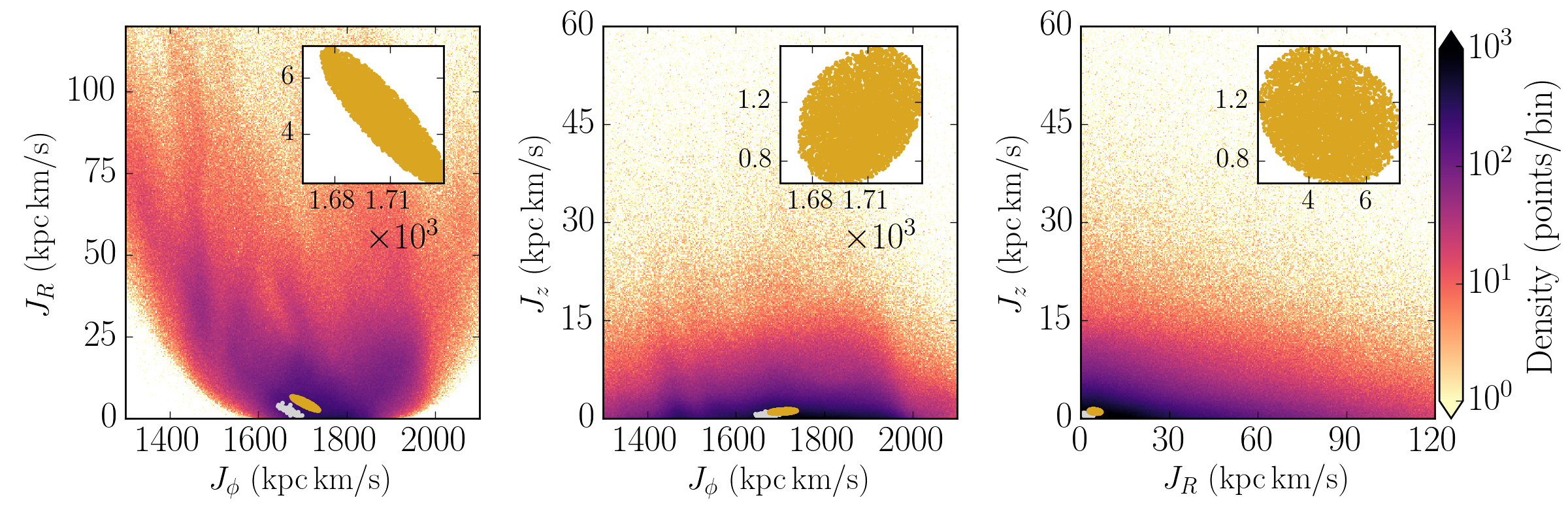}}
\put(-0.02,0){\includegraphics[width=\textwidth]{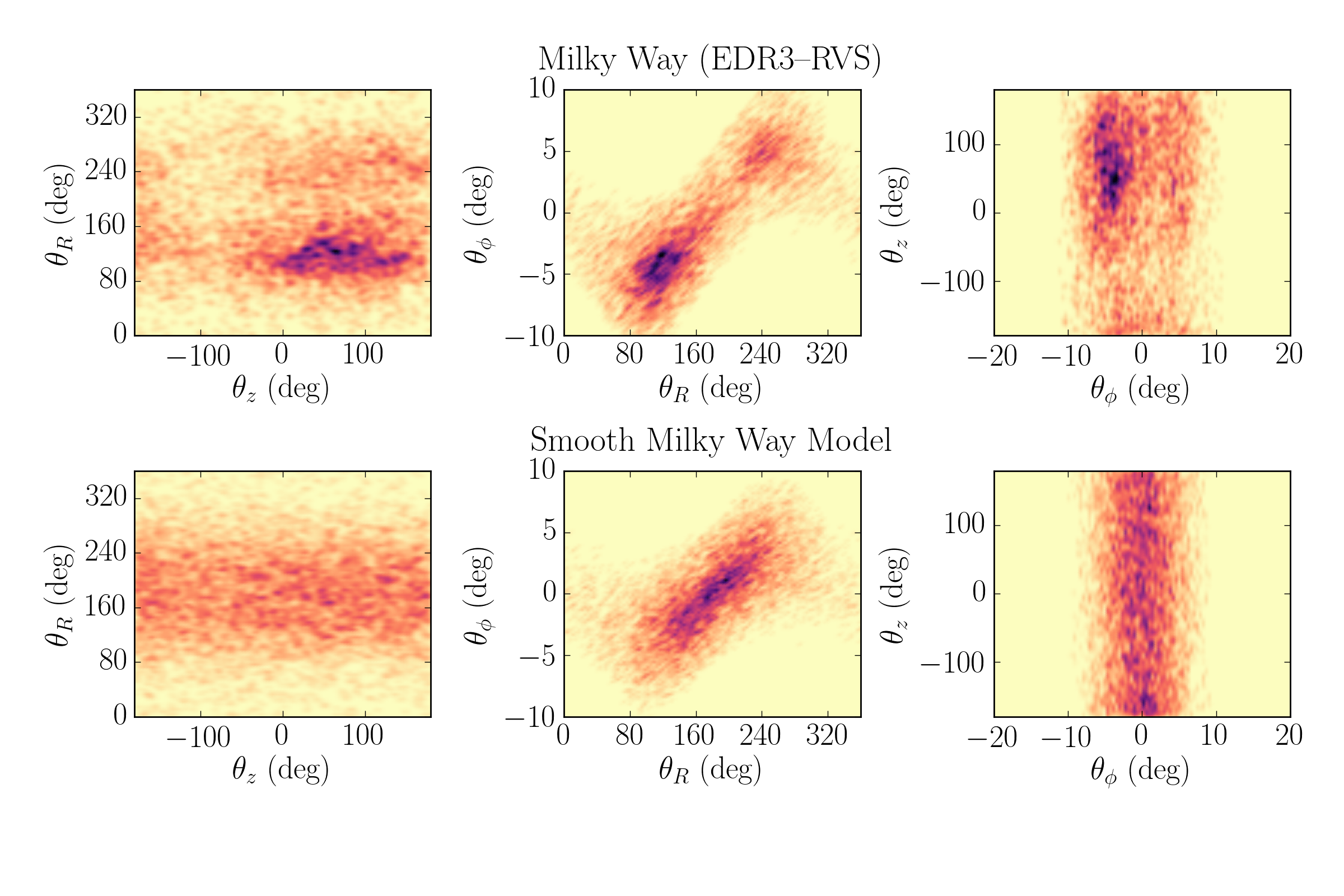}}
\end{picture}
\caption{The orbital phase (angle) distribution in an ``offset'' orbit patch. The top row} shows three projections of the orbital action distribution of the entire Gaia EDR3-RVS, $p({\bf J})~=~p(J_{R}, J_{z}, J_{\phi})$. The orbit patch (golden ellipse) whose angle distribution we illustrate here was chosen to be offset by $3\sigma$ from that of \textit{FoF G12} (grey points).  The middle row shows the 2D projected kernel density maps of the orbital phase (or angle) distribution $p(\theta_{R}, \theta_{z}, \theta_{\phi})$ for all stars in an ``offset'' orbit patch, ${\bf J}\approx {\bf J}_{G12}$. The bottom row shows the analogous distribution in the same orbit patch, but with points drawn from a smooth mock catalog.  
The angle distribution of the Gaia EDR3-RVS data in this offset orbit patch is clearly not as smooth as that of the analogous distribution drawn from the smooth mock catalog. There is some large structure present, e.g. located at $-10 <\theta_{z}\, \rm{(deg)} < 150$, $80<\theta_{R}\, \rm{(deg)} < 160$ and $-10<\theta_{\phi}\, \rm{(deg)} < 0$.
But we do not see multiple compact angle-space overdensities (``pearls''), as seen in the orbit patches centered on the FoF groups (see Figures~\ref{fig:ellipsoid_example}, \ref{fig:angle_ellipsoid_example2} and \ref{fig:angle_ellipsoid_example3}).
\label{fig:G3_9actions_rn}
\end{center}
\end{figure*}

There is much structure present in the angle distribution in orbit patches centered on FoF groups. Obviously far more than in the analogous distributions drawn from a smooth mock catalog. We investigate if such rich angle structure is also present in other orbit patches in the real data of the Gaia EDR3-RVS. We do this by choosing \textit{offset orbit patches} as follows: we start with the orbit patch of one of our 55 FoF group and then shift the patchs' center three times the standard deviation $\sigma$ (in $J_{R}, J_{z}, J_{\phi}$). Through this we pick \textquotedblleft offset\textquotedblright orbit patches that are in a comparably dense part of action space, but \emph{not} centered on an FoF group. As two of the actions ($J_R$ and $J_z$) are positive definite quantities, we take the offsets in the positive action direction. An example of the resulting angle distribution and orbit patch is presented in Fig.~\ref{fig:G3_9actions_rn}.
This orbit patch selection illustrates why we are considering a region well outside the original FoF group (in $J_{\phi}-J_{R}$) and even though we are shifting the center 3$\sigma$, there is still a small amount of overlap in $J_{\phi}-J_{z}$. 

In this offset orbit patch of the real data the orbital phase distribution looks noticeably different: it does show substructure, but there are fewer compact overdensities that could be pearls, compared to Figure~\ref{fig:ellipsoid_example}. 
On the other hand, this angle density distribution also differs clearly from that of the mock catalog: we do see one large but very broad overdensity located at approximately $-10 <\theta_{z}\, \rm{(deg)} < 150$ and $80<\theta_{R}\, \rm{(deg)} < 160$. We should not expect to see a completely smooth distribution as by shifting the location of the group we are nevertheless moving the orbit patch into a high density area of the action plot. In contrast to Figs.~\ref{fig:ellipsoid_example},~\ref{fig:angle_ellipsoid_example2} and~\ref{fig:angle_ellipsoid_example3} the angle distribution of stars within the offset orbit patch does not show any distinct orbital-phase overdensity. 

We proceed to apply this method on the remaining 54 groups and show two more example cases for \textit{FoF G19} and \textit{FoF G40} in the Appendix.

\section{Population Properties of Pearls}
\label{sec:results_disc}

\subsection{Pearls on a string: three case studies}

We are now in a position to analyze in more detail the angle distribution of stars within a FoF-selected orbital patch, drawing on the three examples shown above. We do this foremost by cross-matching the pearls with a catalog of known clusters and structures by \citet{2020A&A...633A..99C} (hereafter CGA20), which has a list of members for 1481 clusters. 
We choose this particular catalog because it is has a large collection of clusters and it is an updated version of the Gaia DR2 cluster census. It also provides membership determinations for known clusters that had been missed by previous studies and additionally for recently discovered clusters.
We further quantify the number of pearls in these orbit patches, and how far apart from one another they are. Finally, we compare all this to simple expectations (the mock catalog and offset orbit patch selection). 

\subsubsection{\textit{FoF G19} and its prominent pearls Platais 8 and IC 2602}
\label{case_study1}

\begin{figure*}
\begin{center}
\includegraphics[width=\textwidth]{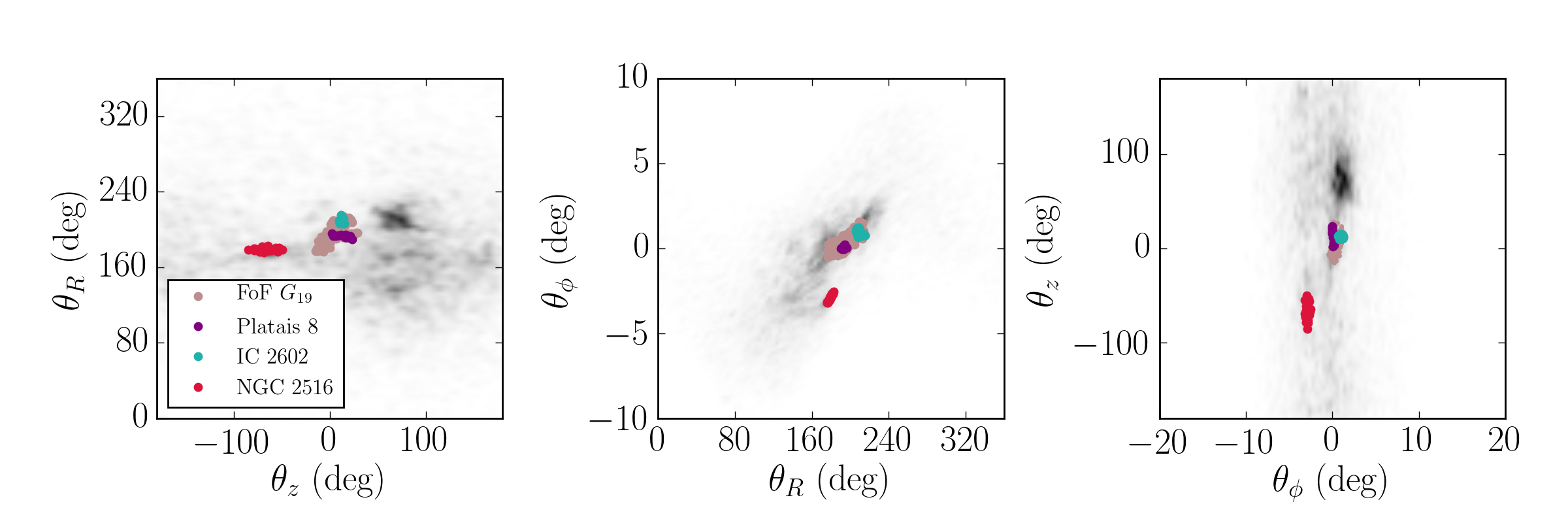}
\includegraphics[width=\textwidth]{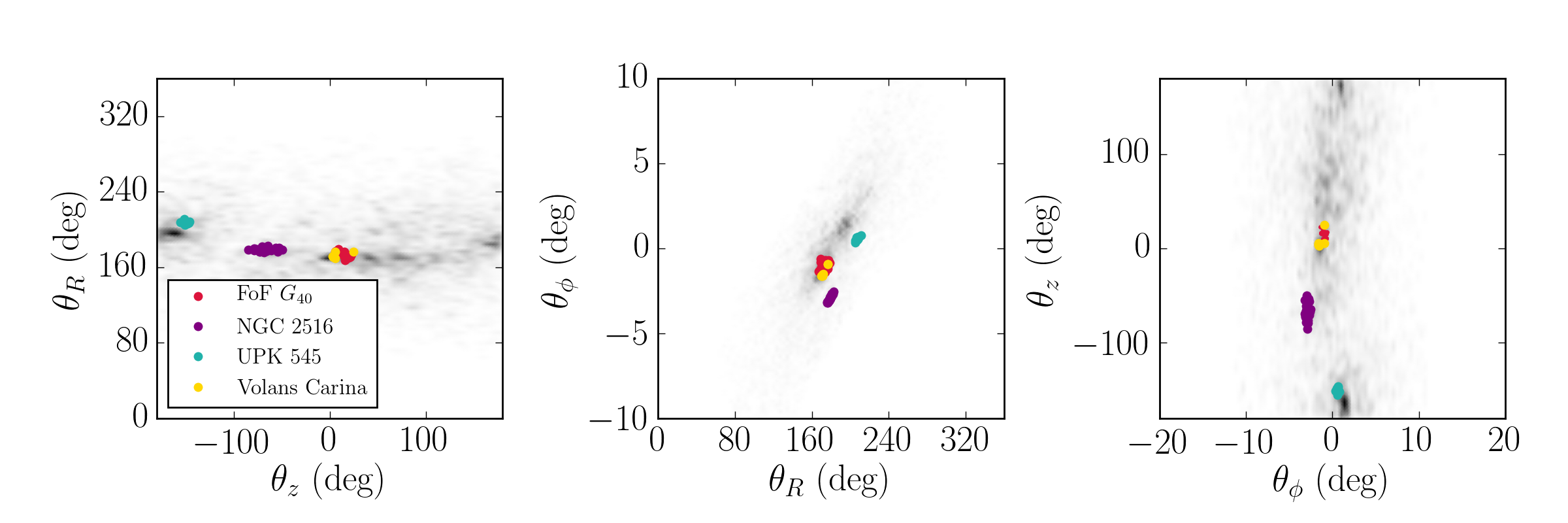}
\includegraphics[width=\textwidth]{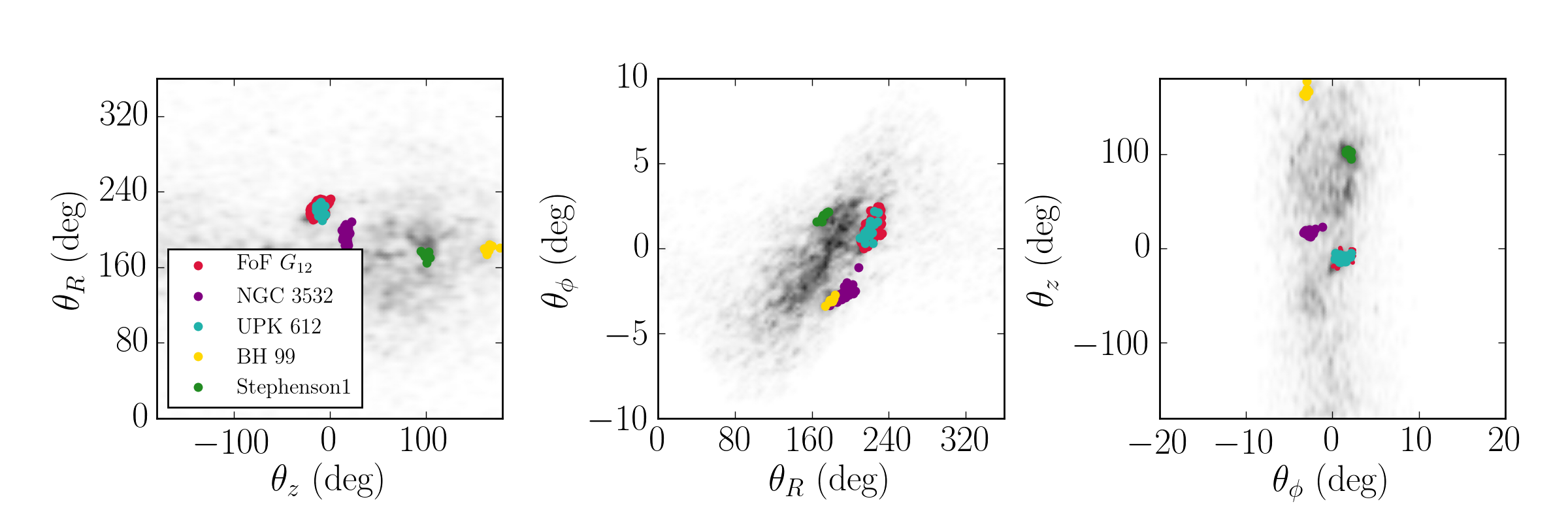}
\caption{Angle distribution of the stars in the \textit{FoF G19}, \textit{FoF G40} and \textit{FoF G12} orbit patches. We cross-match all stars in these FoF orbit patches with members from a catalog of known clusters (CGA20). We highlight these known cluster members with color, along with the members of the orbit-patch defining FoF groups. This Figure illustrates that some pearls are known clusters or associations, others are not.
\newline
\textit{\textbf{Top panel}}: \textit{FoF G19} orbit patch. We find that \textit{FoF G19} is comprised of two known clusters: Platais 8 and IC 2602, shown in purple and blue dots, respectively. The figure shows that \textit{FoF G19} has more members, but none of them are part of a known cluster. We also see two prominent orbital phase overdensities or pearls located at $\theta_{z}$ -- $\theta_{R}\,\rm(deg)\, \sim (-60, 160$) and $(60, 200)$. These are NGC 2516 and UPK 640, respectively.
\newline
\textit{\textbf{Middle panel}}: \textit{FoF G40} orbit patch. We find that some of the most prominent pearls are known clusters: NGC 216 and UPK 545. The group that we identify as \textit{FoF G40} is at the same location than the Volans Carina association. Additionally, we find that some of the stars located just below UPK 545, in the pearl located at $\theta_{R}\,\rm(deg)\, \sim 200$ and $\theta_{z}\,\rm(deg)\, < -100$ belongs to the Pleiades. It is remarkable that all of these pearls seem to be aligned along $\theta_{z}$ and all of them have similar ages, between 90--160 Myr.
\newline
\textit{\textbf{Lower panel}}: \textit{FoF G12} orbit patch. Once more we find that some of the most prominent pearls are known clusters: NGC 3532, UPK 612, BH 99 and Stephenson 1. The group that we identify as \textit{FoF G12} is at the same location as the UPK 612 cluster.}
\label{fig:angles_G1_G2_G3}
\end{center}
\end{figure*}

We start by taking a closer look at \textit{FoF G19}, shown in Fig.~\ref{fig:ellipsoid_example}.
Cross-matching the 231 stars that  are in \textit{FoF G19} and its orbit patch,
we find 46 to also be cluster members in CGA20, mostly members of 
IC 2602 and Platais 8, as shown in the top panel of Fig.~\ref{fig:angles_G1_G2_G3}. These clusters have been identified in the past as OC pairs by \citet{2017A&A...600A.106C}.

It is important to mention that we had already noticed in Fig.~\ref{fig:ellipsoid_example} that (\textit{FoF G19}) appeared bimodal in the vertical action $J_{z}$. In fact this reflects
these clusters, at $J_{z}\sim 0$ we find Platais and above that value we identify IC 2602, which got linked by the FoF approach.

Platais 8 is a cluster located at $\sim$ 135 pc similar to the Pleiades \citep{2018A&A...619A.155S} and IC 2602 is at $d$ = 150--170 pc \citep{2005A&A...438.1163K}. Both clusters are not only on very similar orbits, but also have very similar ages of $\sim$ 60 Myr and 35 Myr, (log(age) = 7.78 and 7.545, \citet{2002A&A...389..871D,2021A&A...645A..84M}), respectively. 

These two clusters are both members of the patch-defining pearl; but there are two more very prominent pearls in this orbit patch, as Fig.~\ref{fig:ellipsoid_example} shows, and many smaller ones. If we now cross-match the entire sample within this orbit patch to the CGA20 catalog, we find many more known clusters. In particular, NGC 2516 and UPK 640 constitute the two most prominent pearls visible in the angle projection of Fig.~\ref{fig:ellipsoid_example}: at $\theta_{z}~-~\theta_{R}\,\rm(deg) \sim (-60, 160)$ and $(60, 200)$, respectively. NGC 2516 is an intermediate-age open cluster of $\sim$ 251 Myr (log(age) = 8.4), located at $d \sim$ 413.8 pc \citep{2021A&A...645A..84M} whereas UPK 640 is among the 207 open star clusters within 1 kpc discovered by \citet{2019JKAS...52..145S} and with an age of 31 Myr (log(age) = 7.5), located at $d \sim$ 177 pc. All of these clusters seem to have very similar young ages, except for NGC 2516 which is slightly older. 

\subsubsection{\textit{FoF G40} and its prominent pearls UPK 545 and NGC 2516}

\textit{FoF G40} is an orbit patch with particularly many pearls, as seen in Fig.~\ref{fig:angle_ellipsoid_example2}. Again, we check which of them correspond to known clusters by cross-matching all stars in this orbit patch to the CGA20 catalog. Also in this case, we find a number of known clusters among the pearls of \textit{FoF G40}; among them are UPK 545 and NGC 2516, which are highlighted in the middle panel of Fig.~\ref{fig:angles_G1_G2_G3}. For UPK 545 the age estimate is $\sim$ 100 Myr \citep{2019JKAS...52..145S}. Furthermore, we cross-match this orbit patch with the catalog of nearby young associations from \citet{2018ApJ...862..138G}, and we find that the action-angle overdensity that defined the \textit{FoF G40} orbit patch is Volans Carina, a 90 Myr old association \citep{2018ApJ...865..136G} (see Fig.~\ref{fig:angles_G1_G2_G3}). 
It is striking that all of these associations are spread widely in $\theta_{z}$, though they have similar ages. Even though there are many more prominent orbital phase overdensities, we were not able to identify the remaining ones located at $\theta_{z} > 20$~deg. However, just below UPK 545, at $\theta_{R}$--$\theta_{z}\,\rm{(deg)} \sim (200,-150)$, we find some stars that have been attributed to the Pleiades, which is $\sim$ 100 Myr old \citep{2018ApJ...863...67G}. Except for NGC 2516, the known clusters that we could identify within the \textit{FoF G40} orbit patch all here have similar ages. It is interesting to note that \citet{2021A&A...645A..84M} provides an age range for many clusters as found in the literature, including NGC 2516. For this particular cluster  \citet{2021A&A...645A..84M}  state a most likely age of 251 Myr, with age limits ranging from 63-299 Myr. 

\subsubsection{\textit{FoF G12} and its prominent pearls UPK 612 and NGC 3532}

A final example, \textit{FoF G12}, is shown in Fig.~\ref{fig:angle_ellipsoid_example3}, with its orbit patch presented in Appendix Fig.~\ref{fig:A_ellipsoid_example3}. 
The angle distribution in Fig.~\ref{fig:angle_ellipsoid_example3} is very clumpy, with many pearls. As before, we cross-match all sample members in the orbit patch with CGA20 and find numerous matches to known clusters, as shown in the lower panel of Fig.~\ref{fig:angles_G1_G2_G3}. The most prominent ones are UPK 612, NGC 3532, BH 99 and Stephenson 1. The UPK 612 cluster stars coincides with the pearl that initially defined \textit{FoF G12}; it is 100 Myr old (log(age) = 8, \citet{2019JKAS...52..145S}). NGC 3532 is 300 Myr old \citep{2019A&A...622A.110F}, Stephenson 1 or $\delta$ Lyrae cluster has an age between 30--100 Myr \citep{2019A&A...630L...8B} and cluster BH 99 is 100 Myr old \citep{2001A&A...379..136C}. Recently, \citet{2019AJ....158..122K} suggested that Stephenson 1 is part of a larger and coeval complex with two more clusters (Gaia 8 and ASCC 100). Additionally, \citet{2019A&A...630L...8B} discovered a star cluster of 100 members (named $\beta$ Lyrae) with indications that
it belongs to a larger old (extinct) star formation complex, including these two known star clusters. Gaia 8 and ASCC 100 are not part of the CGA20 catalog, therefore from the cross-match we only recover Stephenson 1. 

These three case studies that we have shown are typical, with pearls only slightly more abundant than the median case. In most of the cases for our sample of 55 groups we observe very clumpy distributions in angle space, with only a small fraction ($\sim$ 20\%) of FoF-defined orbit patches showing only one distinct pearl. It is also striking that in the three examples we present, the pearls (coinciding also with known clusters) tend to be mostly young, where the oldest one is NGC 3532 (300 Myr). 

\subsubsection{Well-known clusters as pearls}

With the framework we just laid out, we can now put the ``historical'' result into context that the Hyades and Praesepe are two prominent clusters on nearly the same orbit. Indeed both clusters, are among the 55 defining FoF groups (\# 48 Praesepe and \# 49 Hyades, see Table~\ref{tab:long}). Their respective 3$\sigma$ orbit patches overlap, but only modestly. Therefore, the Hyades appear as a pearl of Praesepe (though with fewer than 20 pearl members); the converse is not true, but only because fewer than 10 Praesepe members (with Gaia RVS) fall into the Hyades orbital patch.

But analogous situations are true for pairs of other well-known features. As we detail in 
Appendix~\ref{Appendix_B}, the prominent Coma Berenices cluster (itself FoF group \# 18 in Table~\ref{tab:long}) is a pearl of the extensive Pisces-Eridanus or Meingast 1 stream (which is FoF group \# 14 in Table~\ref{tab:long}). Again, the converse is not true, owing to the fact that Coma Berenice's orbit patch is exceptionally compact.

\subsection{Algorithmically identifying pearls}
\label{quant_pearls}

After this qualitative exposition of the stellar pearls within an orbit patch, we now turn to quantifying this phenomenon. This requires an algorithmic definition of these pearls, and it requires a statistical comparison of the incidence of pearls within an FoF-selected, compared to `offset' orbit patches.


To quantify objectively the number of pearls in an orbit patch, we apply an analogous FoF algorithm to all stars (i.e. stars at all angles) within each of the 55 orbit patches. But of course this FoF analysis is then restricted to the three angle coordinates $p\bigl (\theta_{R}, \theta_{z}, \theta_{\phi}~|~\mathbf{}{J}\bigr )$, as laid out in Eqs. 3 and 4 from \citet{2020MNRAS.495.4098C}. For each orbit patch, we define the linking length as a percentile of the group's pair-to-pair distance distribution in angle space. After some experimentation with different percentiles, we found that a linking length $l_{8^{th}}$, which is the $8^{th}$ percentile of all pairwise angle distances, works best across all groups in picking out ``visually'' distinct, compact pearls. As before, a pearl must have at least three member stars by construction. In each orbit patch the statistics of pearls can then be quantified by the number of pearls $N_{\rm pearls}$ with at least $K$ members. 

The result of this quantitative pearl definition can be seen in Figs.~\ref{fig:pearls_G3_2} and~\ref{fig:pearls_G3_9}, where we illustrate the results of the FoF algorithm to identify the largest orbital-phase overdensities for two examples: the \textit{FoF G40} and \textit{FoF G12} orbit patches, respectively. 
For \textit{FoF G40}, shown in Fig.~\ref{fig:pearls_G3_2}, we find nine rich pearls with $\ge 20$ members. 
For \textit{FoF G12}, shown in Fig.~\ref{fig:pearls_G3_9}, we find 13 pearls with $\ge 10$ members, where we have chosen a lower member-threshold (as there are only 3 pearls with $\ge 20$ members). 

These plots illustrate our basic statistic used for all 55 groups: the number of pearls with $\ge 20$ and $\ge 10$ members. for a given linking length. Subsequently, we calculate the analogous statistic for the 55 corresponding orbit patches in the mock catalog and for the 55 offset orbit patches in the real data. In those latter cases, we use the same linking length $l_{8^{th}}$ as in the FoF-defined orbit patch.

We investigate the ``field star'' contamination of the pearls by identifying the number of sources in the mock catalog that lie within the angle-space region defined by each pearl. We do this by drawing around each member star of a pearl a sphere in angle-space with the radius given by the corresponding linking length. For each pearl we then count the number of stars in the mock catalog that are located inside any of these spheres. For our total set of 238 pearls, we find that half of them have an estimated contamination $<$ 25\%. In other words, pearls typically have a density contrast of 4 over the background in angle-space. About 80\% of the pearls have an estimated contamination $<$ 50\%. 

If group and pearl identification is as powerful as argued here, this significant level of contamination may cause some consternation, as stellar associations have been picked out with lower contamination in the, arguably more intuitive, space of common space velocities. In this context it is important to appreciate that here we are probing a new regime of sparse and dispersed associations (extended over several 100 pc) among stellar associations with relatively few members. Finding them in position or (better) space-velocity space would be all but hopeless for most low-contrast pearls. Here, we show that we can identify them in orbit-space, though with some contamination. Pearls that are clear (low-contamination) stand-outs in velocity space, are also clean in orbit space.

\subsection{On-sky distribution of pearls on the same string}

\begin{figure*}

\setlength{\unitlength}{\textwidth}
\begin{center}
\begin{picture}(1,1)
\put(0,0.65){\includegraphics[width=\textwidth]{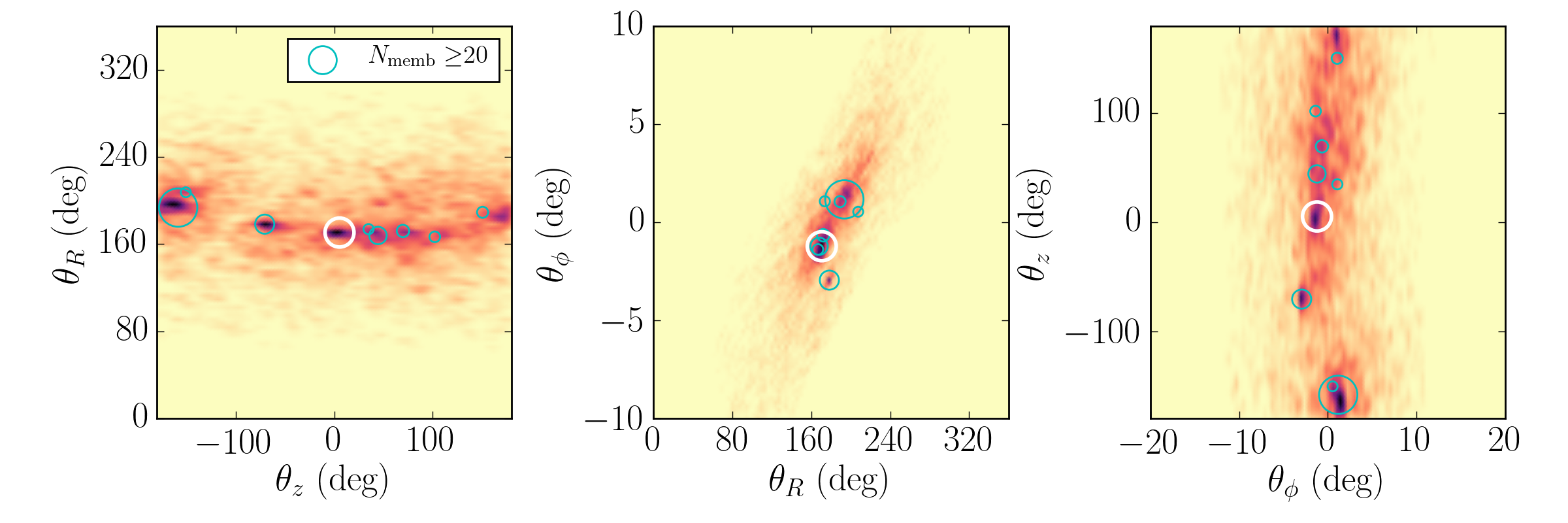}}
\put(0.05,0.32){\includegraphics[width=0.5\textwidth]{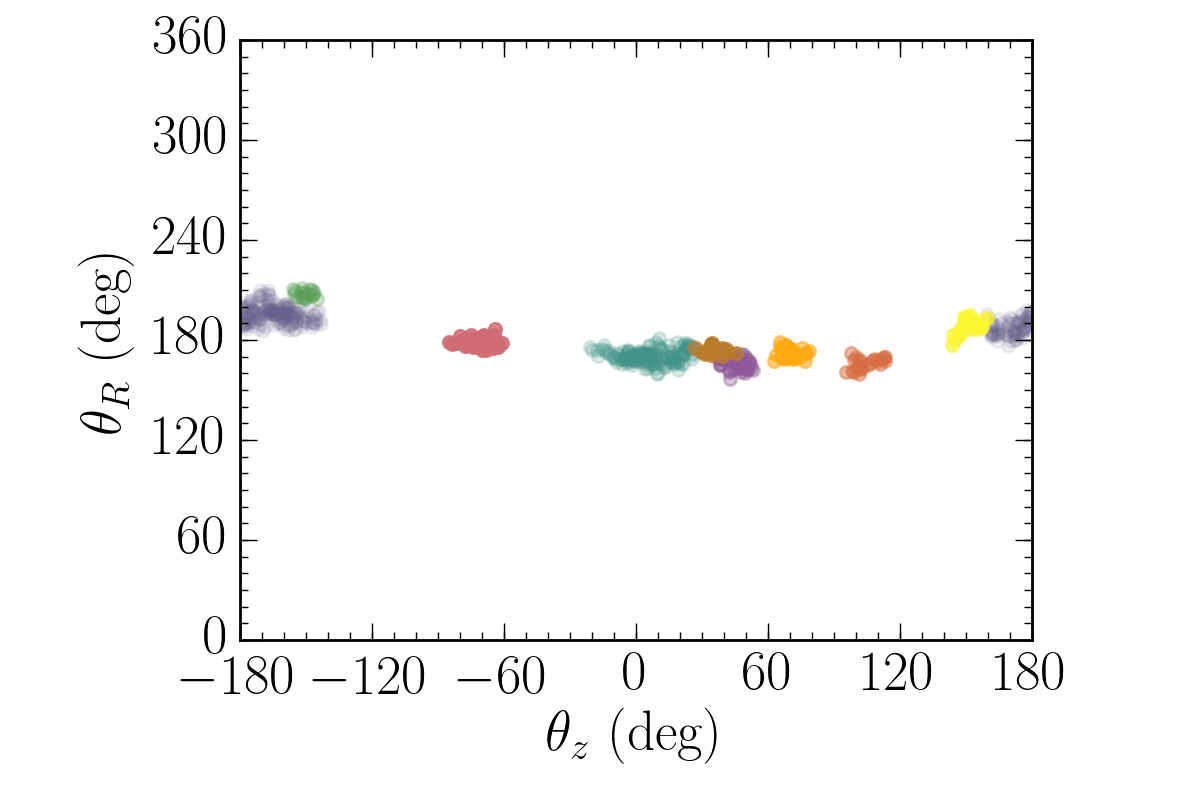}}
\put(0.5,0.32){\includegraphics[width=0.5\textwidth]{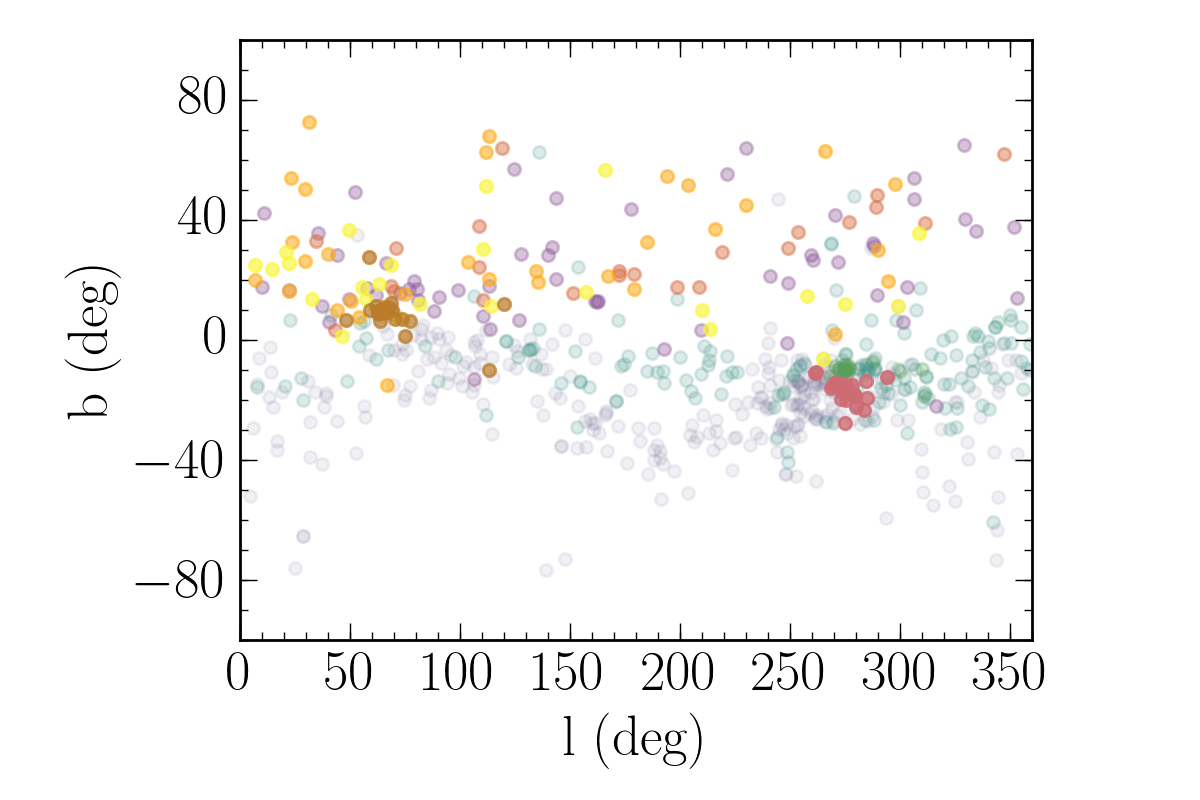}}
\put(0,0){\includegraphics[width=\textwidth]{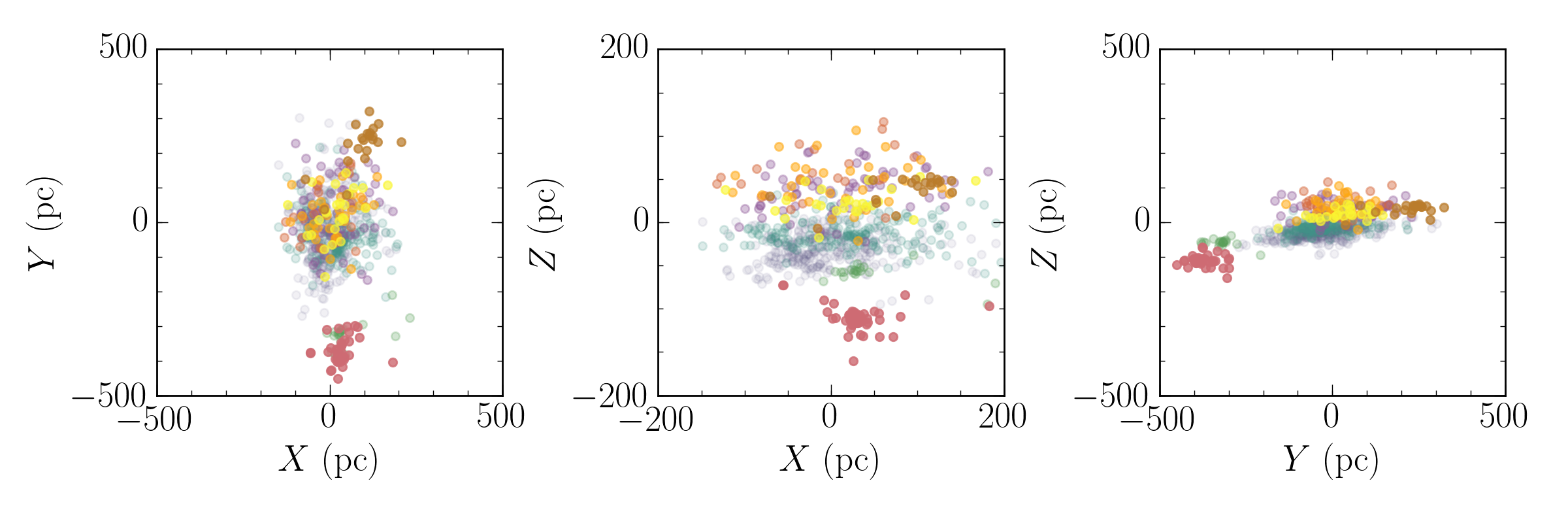}}
\end{picture}

\caption{Algorithmic identification of the orbital-phase overdensities, i.e. pearls, for the \textit{FoF G40} orbit patch, using a FoF algorithm restricted to 3D angle space, with a linking length that is 8\% of the median pair separation.\newline
\textit{\textbf{Top panel}}: We encircle each identified pearl and the size of the circle scales with the size (number of members) of the pearls. Here we show the 9 largest pearls that the FoF finds with a minimum number of members of 20, illustrating the most prominent ones. The location of the original group (\textit{FoF G40}) is illustrated with a white circle. There is a pearl located at $\theta_{z}$ -- $\theta_{R}\,\rm(deg)\, \sim (150, 180)$ that because of the periodicity of the angles is split in $\theta_{z}$ and it is actually part of the same pearl located at  $\theta_{z}$ -- $\theta_{R}\,\rm(deg)\, \sim (-150, 180)$. This can be seen more clearly in the middle panel, left side.
\newline
\textit{\textbf{Middle panel left}}: Angle distribution of these same pearls but now color coded to distinguish them individually. The most prominent and largest pearl (in pale blue dots) is split in $\theta_{z}$ (because of the periodicity of the angles). \newline
\textit{\textbf{Middle panel right}}: Distribution of these pearls in \textit{l,b}, where we see that some of them (pink and brown for example) are clumped in Galactic longitude and latitude. However, the largest pearls (blue and green) are spread all over the sky. The blue pearl especially extends all over $l$ and seems to oscillate in $b$.\newline
\textit{\textbf{Lower panel}}: Position in rectangular Galactic coordinates $X,Y,Z$ of these pearls. The $Z$ coordinate is positive pointing towards the North Galactic pole, $X$ increases in the direction of the Galactic center, the Sun is located at $(0,0,0)$ and $Y$ is pointing in the direction of rotation of the Galaxy. We notice that some pearls (pink and green) are several hundred parsec away from the rest.}
\label{fig:pearls_G3_2}
\end{center}
\end{figure*}

\begin{figure*}
\setlength{\unitlength}{\textwidth}
\begin{center}
\begin{picture}(1,1)
\put(0,0.65){\includegraphics[width=\textwidth]{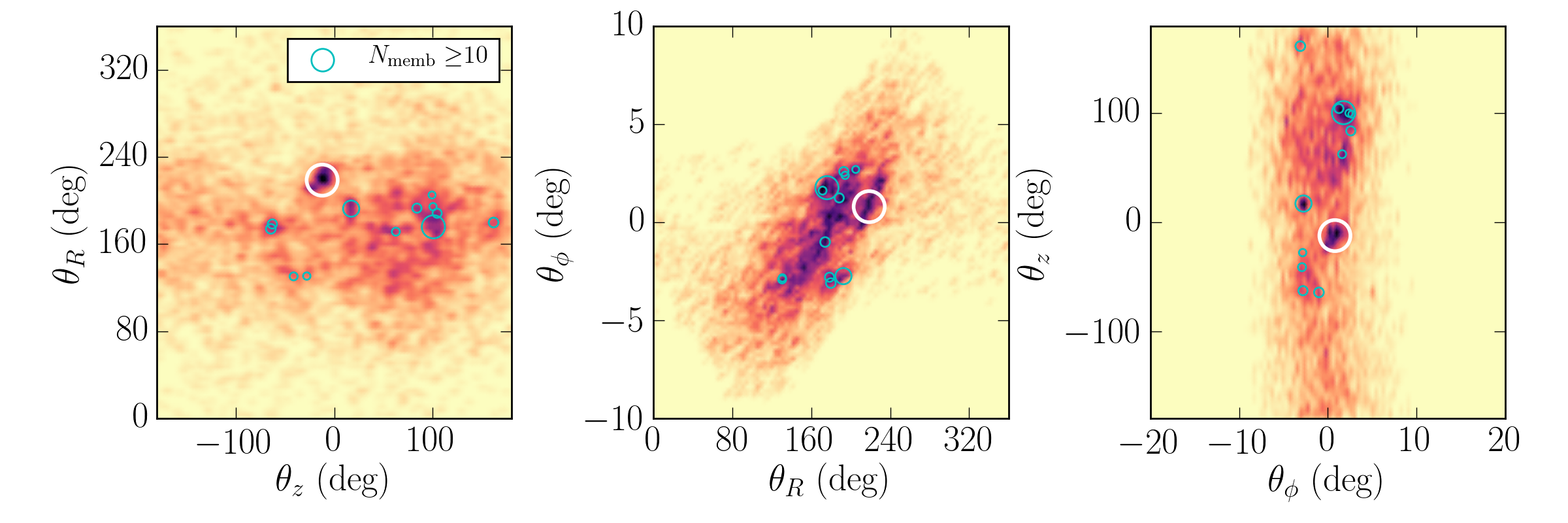}}
\put(0.05,0.32){\includegraphics[width=0.5\textwidth]{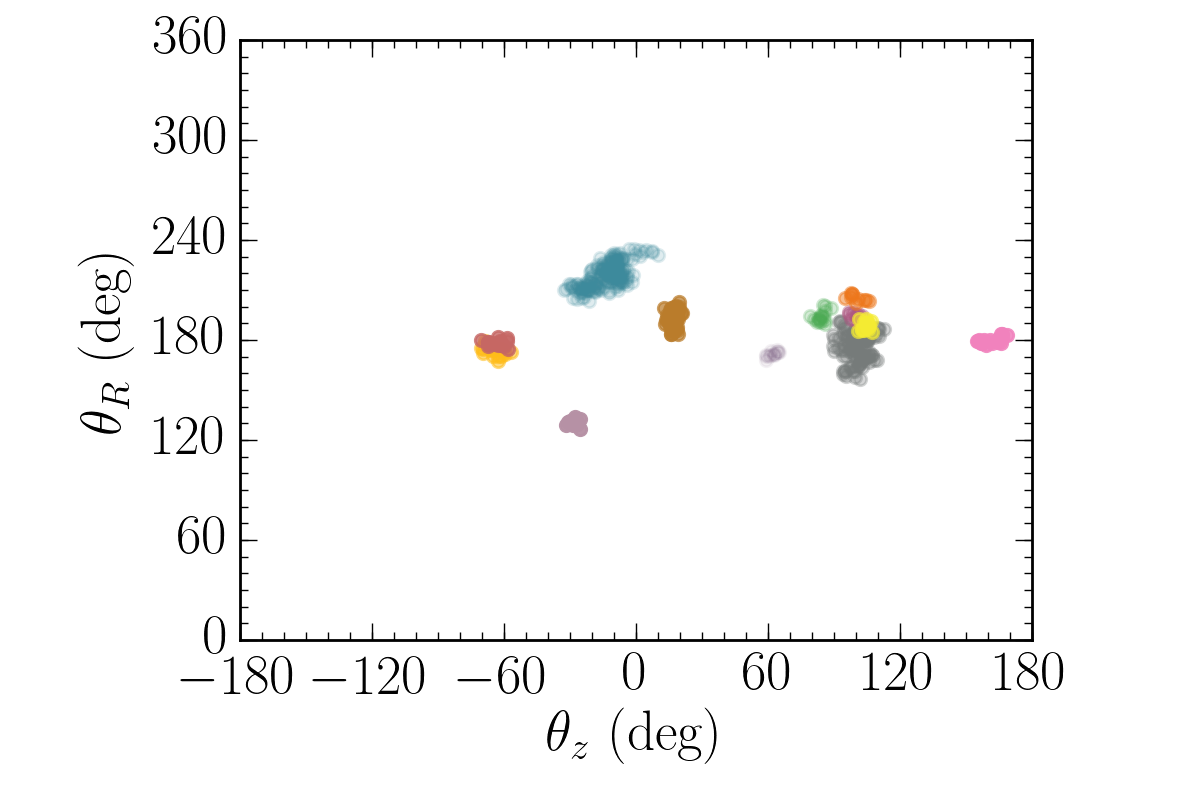}}
\put(0.5,0.32){\includegraphics[width=0.5\textwidth]{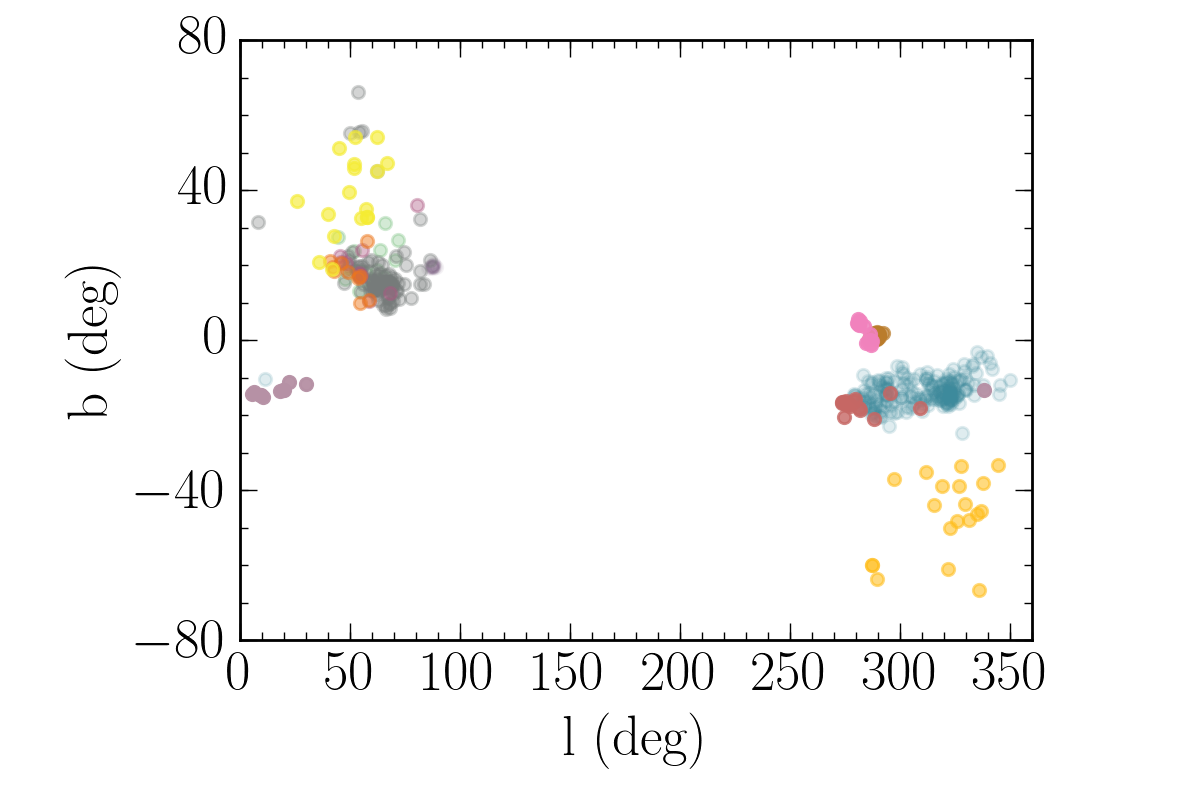}}
\put(0,0){\includegraphics[width=\textwidth]{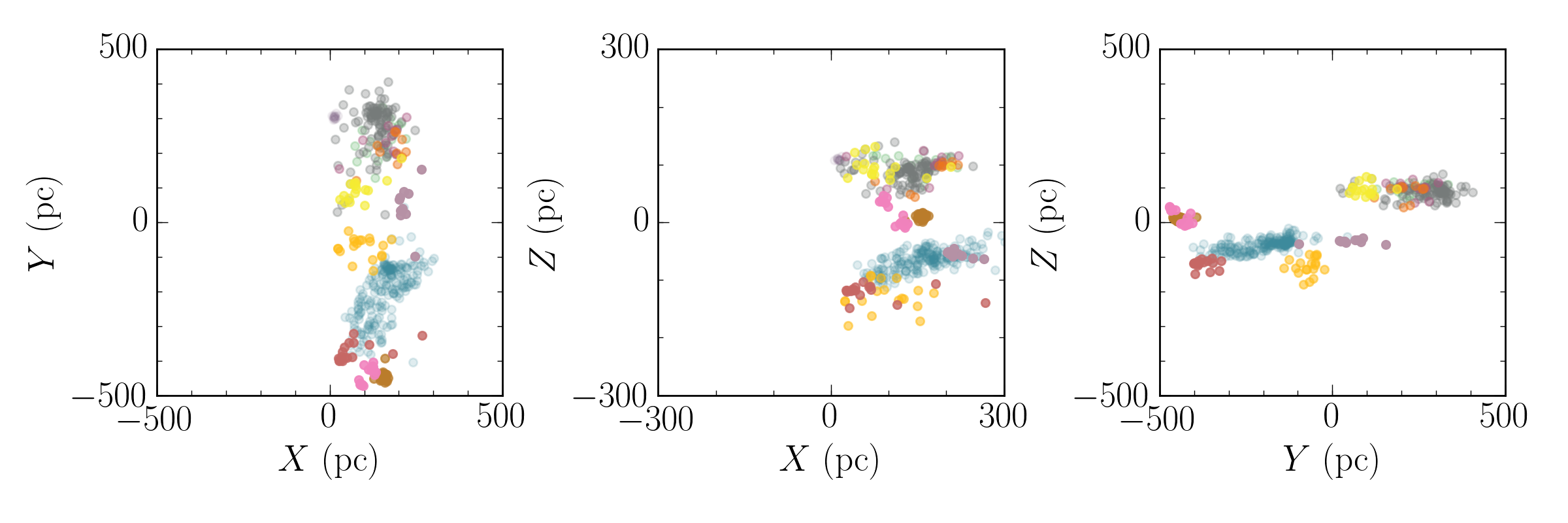}}
\end{picture}

\caption{Algorithmic identification of the orbital-phase overdensities, i.e. pearls, for the \textit{FoF G12} orbit patch, using a FoF algorithm restricted to 3D angle space, with a linking length that is 8\% of the median pair separation.
\newline
\textit{\textbf{Top panel}}: We encircle each identified pearl and the size of the circle scales with the size (number of members) of the FoF group. In this case we chose a minimum number of 10 FoF members, yielding 13 pearls. With a minimum of 20 members the FoF still finds 3 pearls. The location of the original group (\textit{FoF G12}) is illustrated with a white circle. 
\newline
\textit{\textbf{Middle panel left}}: Angle distribution of the same pearls now color coding each one to distinguish them individually.\newline
 \textit{\textbf{Middle panel right}}: Distribution of these pearls in \textit{l,b}, where we see that while some of them (pink and brown) are confined in Galactic longitude and latitude ($l,b$) there are some that are spread in the sky (for example the yellow and orange pearls).\newline
 \textit{\textbf{Lower panel}}: Position in rectangular Galactic coordinates $X,Y,Z$ of the same pearls. The $Z$ coordinate is positive pointing towards the North Galactic pole, $X$ increases in the direction of the Galactic center, the Sun is located at $(0,0,0)$ and $Y$ is pointing in the direction of rotation of the Galaxy. We notice that these pearls are separated by several hundred parsec.}
\label{fig:pearls_G3_9}
\end{center}
\end{figure*}

Now that we have properly identified the orbital-phase overdensities we want to explore to which extent they should be characterized as clusters, associations, or streams. And we want to explore how widely they are distributed across the accessible angle-space. We do this by looking at distribution of pearls and pearl members on the sky. The middle left panel of Fig.~\ref{fig:pearls_G3_2} shows the the algorithmically found pearls, the members of each in a distinct color. By construction, they are compact in angle-space, with only the most prominent and largest pearl seemingly split in $\theta_{z}$, due to the periodicity of the angles. The distribution of the pearl members on the sky is shown in the middle right panel of Fig.~\ref{fig:pearls_G3_2}, in Galactic longitude and latitude ($l,b$). Some pearls are tightly clustered in ($l,b$), whereas other pearls show very extended distributions: they are spread all across the sky. Interestingly, the largest pearl (in pale blue) does not only extend across all $l$, but also oscillates up and down in $b$. This range of on-sky appearances is explained through the lower panel of Fig.~\ref{fig:pearls_G3_2}, which shows the position of the pearls in Galactic Cartesian coordinates $(X,Y,Z)$ centered on the Sun. It shows that many pearls span several hundred parsec, an extent comparable to their mutual separation (in X,Y,Z). And of course, pearls that are nearly centered on the Sun's location, will appear all around us. In Fig.~\ref{fig:pearls_G3_9} we illustrate the sky distribution for \textit{FoF G12}. While some of the pearls in \textit{FoF G12} are spread in $(l,b)$, but not to the same extent as in Fig.~\ref{fig:pearls_G3_2}. 
This is presumably because the orbit generating \textit{FoF G12} does not cross the Solar locus.
Clearly, most pearls are more extended than clusters, or even associations. To which extent we are witnessing associations in the process of dissolution is beyond the scope of the present paper, but will be analyzed in forthcoming work.

\subsection{The statistics of pearls}
\label{sec:frac_pearls}

\begin{figure*}
\centering
\hspace*{-0.2cm}\includegraphics[width=.5\textwidth]{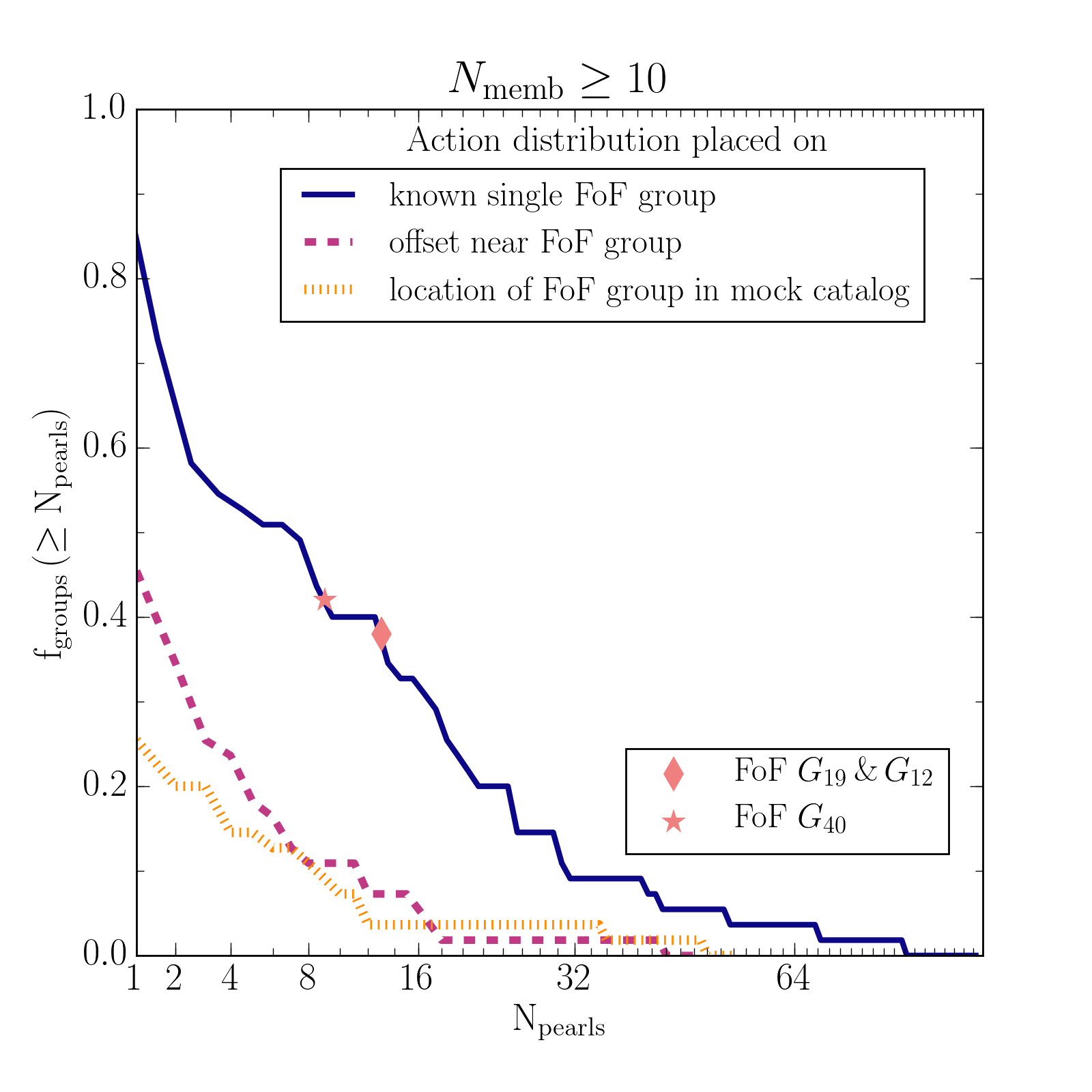}\hspace*{-0.2cm}
\hspace*{-0.2cm}\includegraphics[width=.5\textwidth]{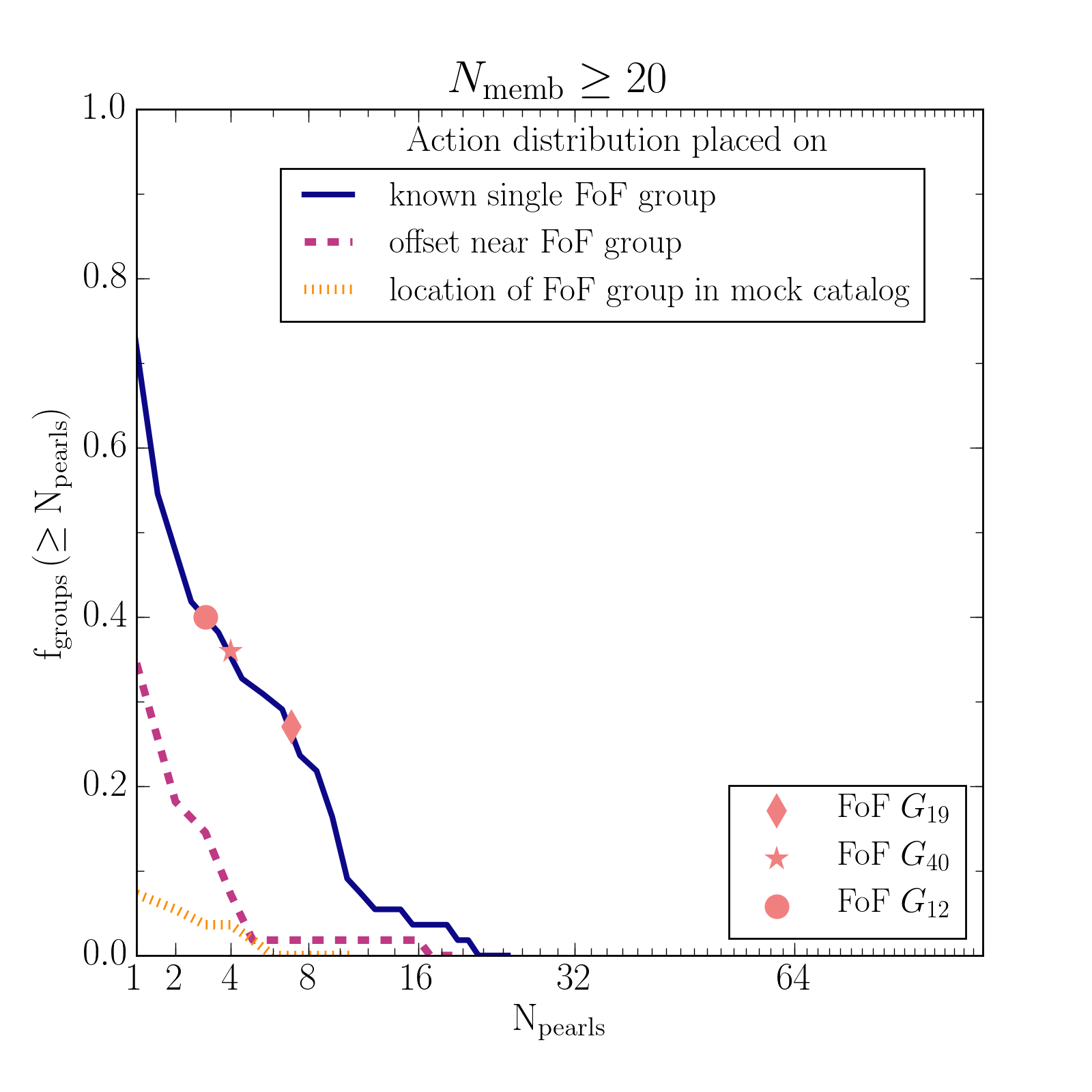}\hspace*{-0.2cm}\\
\caption{Fraction of groups as a function of the number of pearls for the orbit patches in the 55 groups that we find with the FoF algorithm in angle space. We consider two cases: pearls with a minimum number of 10 (left) and 20 members (right). Our orbit patch selection for the FoF groups is shown with a blue solid line, the mock orbit patch with a purple dashed line and the offset orbit patch with an orange dotted line. The \textquoteleft patch-defining pearl\textquoteright\,in the known single FoF group (blue solid line) is subtracted in this figure. 
\newline
\textit{\textbf{Left}}: With the orbit patch selection for the FoF groups, all of them have at least one pearl and 0.7 have 3 pearls. For the offset orbit patch this fraction is $\sim$ 0.4 and for the mock case it drops to $\sim$ 0.2 of the groups having one pearl. Less than 0.1 of the groups in both the offset and mock orbit patch have 5 pearls, whereas for the known FoF groups this fraction is almost 0.6. 
\newline
\textit{\textbf{Right}}: In this case we find almost one pearl ($\sim 0.95$) in all of the known FoF groups orbit-pace patches. For the offset orbit patches this fraction is $\sim$ 0.38 and for the mock case this drops to almost 0. For the known FoF groups orbit-pace patches $\sim$ 0.6 of them have 2 pearls. Most of the groups that we find have others on similar orbits. } 
\label{fig:n_pearls}
\end{figure*}

We now have an algorithmic way to quantify the statistics of pearls in each of the 55 orbit patches, $N_{\rm pearls}(\ge K_{\rm member})$, using the FoF algorithm in angle space. On this basis we can investigate how common it is to find more than one pearl within an orbit patch. We do this for the 55 FoF-selected orbit patches, and compare this to the statistics of the the mock and the offset orbit patches. We consider two cases, pearls with a minimum of 10 and 20 members. These statistics are illustrated in Fig.~\ref{fig:n_pearls}, where we show the fraction of orbit patches among our 55 FoF groups that show at least $N_{\rm pearls}(\ge 10)$ and $N_{\rm pearls}(\ge 20)$. 
In this figure we subtract the \textquoteleft patch-defining pearl\textquoteright\,in the FoF-selected orbit patches so we can make an unbiased comparison to the mock and offset orbit patches. We find that 50\% of FoF-selected orbit patches have 7 additional pearls of $\ge 10$ members, and 20\% of these groups show even 25 additional pearls of $\ge 10$ members. 
When considering ``rich'' pearls of at least 20 members, we find that 50\% of the FoF-selected orbit patches show two additional rich pearls, while 20\% show eight additional rich pearls. These frequencies of pearls are markedly different from the analogous statistics derived for the orbit patches in the mock catalog and the offset orbit patches: there, we find
that $\ge$50\% contain no pearls of 20, or even 10 members. For the mock catalog, barely 20\% show at least one pearl of $\ge 10$ members; while only 7\% show one (but not more) pearl of $\ge 20$ members. For the offset orbit patches in the real data, the analogous numbers are: 20\% show 4 pearls of $\ge 10$ members, and two pearls of $\ge 20$ members (see Fig.~\ref{fig:n_pearls}). 

Put differently, the majority of other orbit patches shows no pearls of $\ge 10$ members, the vast majority of other orbit patches shows no pearls of $\ge 20$ members. By contrast, most orbit patches centered on one known pearl, are rich in other pearls (beyond the one that defined the orbit patch). An interesting subset are the 8 (15\%) of FoF orbit patches that only show the defining pearl: orbits with an isolated pearl. We show an example in Appendix~\ref{sec:isolated_groups}). There the only pearl is Platais 3, a known cluster with a (literature) age of nearly 1 Gyr (log(age)=8.92, \citet{2018A&A...615A..12Y}). Similarly, a number of the other isolated pearls are known clusters of ages well above $t_{dyn}\sim 200$~Myrs. Qualitatively it is unsurprising that clusters that are ``old'' and bound are more likely found without discernable birth cousins in pearls on the same orbit. Furthermore, older FoF groups must be particularly compact in action space, if they are to remain also compact in angle space. And of course more compact orbit patches will contain far fewer other stars, boosting their chances to appear as isolated pearls. We will pursue the question of consistent age-dating of pearls, and of its implication in a forthcoming paper. 

Fig.~\ref{fig:n_pearls} also puts our three case studies into context: in both the cumulative statistic for pearls of 10 and of 20 members they lie around the 40$^{th}$ percentile: they are typical, with pearls only a little bit more abundant than the median case.

We now address the question of whether the set of all pearls in all orbit patches are actually disjoint sets of stars. Per construction, a star member of a pearl in one orbit patch can also be part of a pearl in another orbit patch. 
This is because the orbit patches in this context are three times larger than the standard deviation in orbital actions in the FoF group that spawned this orbit patch.
To explore this, we have asked of all stars that are members of at least one pearl, whether they are also members in one or more additional pearls in other patches.  We find that for pearl members stars the large majority of them ($\sim$ 75\%) are members of only one pearl. However,  about 20\% are members of 2 or 3 pearls, and about 5\% are members of 4 or more pearls.  This is true both when considering pearls with at least 10 and at least 20 members.

This 25\% overlap among pearls stars belonging to different groups appears to be another manifestation of how clustered orbit-space is. Technically, our 55 FoF groups are -- by construction -- indeed fully distinct and disjoint. Yet, their surrounding 3-$\sigma$ action-angle space ellipsoids, which we use
to search for pearls, do indeed start to overlap among groups in about 25\% of cases. As the total fraction of the pertinent action-angle space that is filled by the initial 55 FoF groups is small, this shows how clustered orbit space appears to be. This clustering, presumably a reflection of the structured ISM, deserves further quantification in future work. 

\section{Summary and Discussion}
\label{sec:summary}

In this paper we provide the first quantification of the correlation of recent Galactic disk star formation in \emph{orbit} and \emph{orbital phase} space.  The goal was to provide an empirical basis for understanding how the large-scale filamentary structure of the cold ISM \citep{Alves2020} might  be reflected in the orbit distribution of the stars that have formed from it (within $t_{\rm age}\le t_{\rm dyn}(R_\odot )\approx 200$~Myrs). We were particularly interested in the large-scale ($\gtrsim 100$~pc) correlations among distinct clusters, associations or other aggregates of presumably young stars.  

The work presented here builds on earlier quantifications of large-scale correlations among such clusters in configuration-space ($\vec{x},\vec{v}$), e.g. by \citet{2017A&A...600A.106C} and \citet{2019AJ....158..122K}. And it builds on the longstanding anecdotal evidence that different clusters, widely separated in the sky, can be on very similar obits \citep[e.g.][]{1959Obs....79..143E}.

There are three new elements we could bring to the issue here: First, the use of action-angle coordinates ({\bf J},$\theta$), which are the canonical coordinates for orbit space. In particular, they offer a neat separation of the 6D coordinates in orbit (action), and orbital phase (angle); and it turns out that $p(\theta~|~{\bf J})$, the distribution of angles in a given obit-patch is a fruitful way to look at the problem at hand. Second, we do not rely on traditional catalogs of distinct clusters, associations, etc, but find overdensities of stars in actions-angle space algorithmically using a friend-of-friends algorithm (FoF) in ({\bf J},$\theta$)-space applied to all stars with radial velocities within Gaia EDR3 and distances $< 800$~pc, as laid out in our preceding work \citep{2020MNRAS.495.4098C}. Third, the Gaia EDR3 data with radial velocities (EDR3-RVS) is a superb data set to explore the 6D stellar phase-space distribution in the 1~kpc around the Sun.

Following \citet{2020MNRAS.495.4098C} we start by finding the 55 most prominent clusterings of EDR3-RVS member stars in action angle space $({\bf J},\theta)_{i=1,55}$ using a 6D FoF algorithm. Each of these defines an ``orbit patch'', and ellipsoid in action space around ${\bf J}_i$. We then consider \emph{all} stars within this orbit patch, i.e. with actions $\approx {\bf J}_i$, and look at their angle distribution, 
$p(\theta~|~{\bf J}_i)$. By construction (via a 6D overdensity), the distribution $p(\theta~|~{\bf J}_i)$ must have at least one compact overdensity in angle space. However, within most of these orbit patches we find many more compact angle (or orbital-phase) overdensities; and we quantify these angle-space overdensities by a 3D FoF algorithm (with a linking length of 8\% of the sample median separation). We find that these 55 obit patches typically contain 8 more overdensities with $\ge 10$ EDR3-RVS members; they have 20 more angle-space overdensities in 20\% of the cases, strung along the same orbit but at different orbital phases.
For nomenclature convenience, we dub these angle-space overdensities \textquoteleft pearls\textquoteright. Detailed investigation of these pearls among three of these obit patches reveals that these pearls often are well-known clusters or associations; but often, they are previously unrecognized associations, spread out in configuration-space, yet compact in action-angle space. 

In the three sample FoF groups where a few pearls coincide with clusters of robust (literature) ages, these tend to be young $t_{\rm age}<200 \sim$Myrs. Clearly, a systematic approach to determining ages of pearls, and checking whether they are mono-age populations, is crucial to put these findings into a broader context. For instance, it could clarify whether really most of them are young ($<$ few 100~Myrs) and dispersing but not yet phase-mixed. In at least one case (see Appendix B) we could show with literature data that not all pearls on the same string will be approximately co-eval; this means that being pearls on the same string does not necessarily or always imply any form of 'common birth origin'. Unfortunately, our 6D sample only contains stars over the very limited range on T$_{eff}$ for which RVS information is available \citep{Rybizki2021}; color magnitude diagrams for those stars exclude the turn-off region of young populations and do not permit meaningful age constraints. We therefore defer a comprehensive age estimate of all pearls to a forth-coming paper.

We then compare the incidence of such pearls found in the 55 orbit patches centered on one prominent pearl with their incidence in two alternate control sets of orbit patches. One set of obit patches centered on the same ${\bf J}_{i=1,55}$, but with the member stars drawn from a smooth mock catalog \citep{2018PASP..130g4101R}. Another one, with orbit patches that are offset from the actual ${\bf J}_{i=1,55}$, and with the members drawn from the actual EDR3-RVS. In both cases, we find a dramatically lower incidence of pearls.

This implies that recent star-formation in the Galactic disk is strongly clustered towards a modest subset of particular obits, presumably the orbits on which the cold ISM was moving when giving birth to these stars. This opens up new avenues of studying the dynamics of the ISM, as it is generally impossible to get 6D phase-space coordinates for it, except for the relatively sparse set of masers \citep{2014ApJ...783..130R}.
To understand the orbits and positions on which stars are actually born, it would also be good to calculate their obits back towards their birth position. This requires testing whether pearls are mono-age populations, and determining their ages systematically. In this paper, we have only shown that pearls that are well-studied clusters tend to be young ($\lesssim 100$~Myrs). A more systematic exposition of this issue will be part of a forthcoming paper. Then we should be in a position to see whether the orbits on which stars are preferentially born are a stochastic sub-sample of disk-like orbits, or are discernible ``special'' in the overall disk dynamics relating, for example, to orbit resonances.

But for now, we could quantify the extent to which young stars are clustered in orbit space: there is a subset on which distinct aggregates of young stars are common, albeit at widely separate phases. Indeed, stars in the Galactic disk seem to form like pearls on a set of strings.


\begin{acknowledgments}
We thank the anonymous referee for their exceptionally thorough and constructive report that greatly improved the paper.
We thank the MPIA MW group and Greg Green for providing very useful comments. J.C. acknowledges support from the International Max Planck Research School for Astronomy and Cosmic Physics at Heidelberg University (IMPRS-HD). H.W.R. received support from the European Research Council under the European Union's Seventh Framework Programme (FP 7) ERC Grant Agreement n. [321035].
This work has made use of data from the European Space Agency (ESA) mission Gaia\footnote{\url{http://www.cosmos.esa.int/gaia}}, processed by the Gaia Data Processing and Analysis Consortium (DPAC)\footnote{\url{http://www.cosmos.esa.int/web/gaia/dpac/consortium}}. Funding for the DPAC has been provided by national institutions, in particular the institutions participating in the Gaia Multilateral Agreement. 
\end{acknowledgments}

%






\appendix
\section{3D Action ellipsoid continued}
\begin{figure*}
\begin{center}
\includegraphics[width=\textwidth]{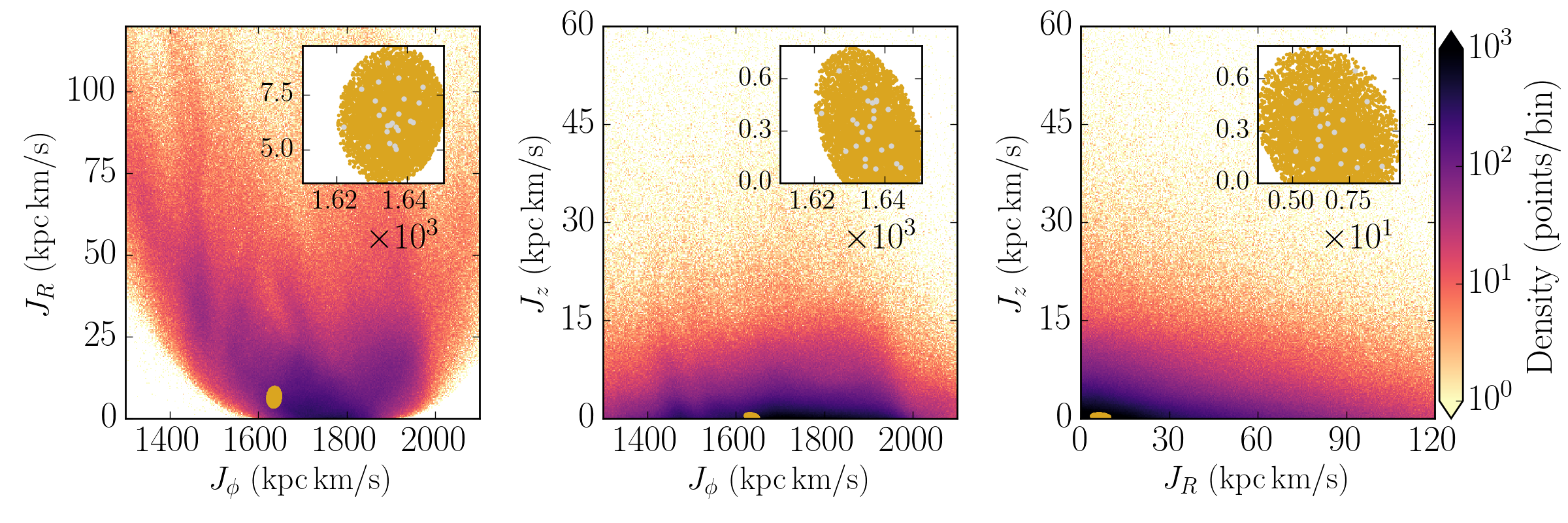}
\caption{Action distribution ($J_{R}, J_{z}, J_{\phi}$) for the entire GEDR3-RVS selection and the \textit{FoF G40} orbit patch. With yellow dots we show the orbit patch selection and in grey dots we show the group members. The upper right inset in each panel shows a zoom-in into the orbit patch to highlight clearly its size and also how well this method encloses the group.}
\label{fig:A_ellipsoid_example2}
\end{center}
\end{figure*}

\begin{figure*}
\begin{center}
\includegraphics[width=\textwidth]{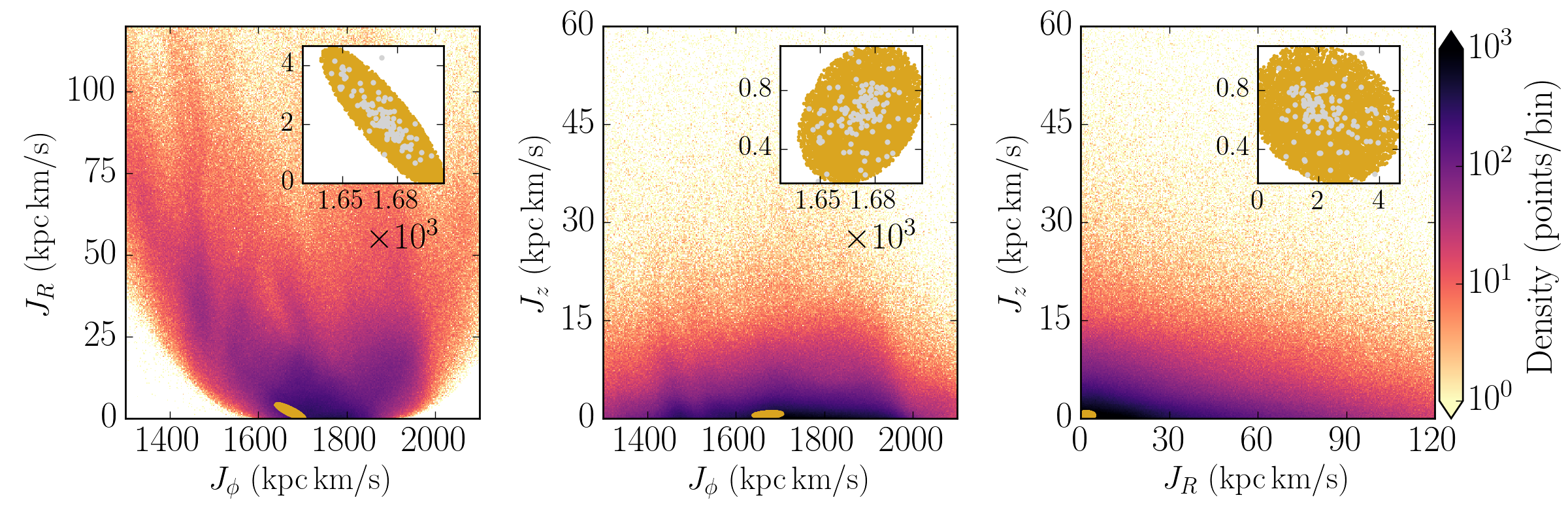}
\caption{Action distribution ($J_{R}, J_{z}, J_{\phi}$) for the entire GEDR3-RVS selection and the \textit{FoF G12} orbit patch. With yellow dots we show the orbit patch selection and in grey dots we show the group members. The upper right inset in each panel shows a zoom-in into the orbit patch to highlight clearly its size and also how well this method encloses the group.}
\label{fig:A_ellipsoid_example3}
\end{center}
\end{figure*}

Here we show the orbit patches for the remaining two example groups: \textit{FoF G40} and \textit{FoF G12}. We find 24 members for \textit{FoF G40} and 122 for \textit{FoF G12}. The corresponding orbit patches are presented in Figs.~\ref{fig:A_ellipsoid_example2} and~\ref{fig:A_ellipsoid_example3}. 

\section{Case study: Coma Berenices as Pearl of Pisces Eridanus}
\label{sec:Appendix_B}

Another interesting case study is the Pisces Eridanus or Meingast 1 stream. We identify one of our FoF groups as the stream by inspecting its location in position ($\alpha, \delta$) and then cross-matching its stars with the candidates and members of this stream presented in \citet{2019AJ....158...77C}. Pisces Eridanus is a recently discovered stream, and it has been investigated only by a few studies, for example: \citet{2019A&A...622L..13M}, \citet{2019AJ....158...77C} and more recently by \citet{2020MNRAS.496.2422H,2020A&A...639A..64R}. \citet{2019A&A...622L..13M} first discovered the stream, and besides a detailed kinematic study, they also estimated and age of $\sim$ 1 Gyr through isochrone fitting, using the PARSEC isochrone library \citep{2012MNRAS.427..127B}. However, later studies \citep{2019AJ....158...77C,2020A&A...639A..64R} concluded that the stream has a similar age as the Pleiades, $\approx$ 120 Myr.

\begin{figure*}
\setlength{\unitlength}{\textwidth}
\begin{center}
\begin{picture}(1,1)
\put(0,0.65){\includegraphics[width=\textwidth]{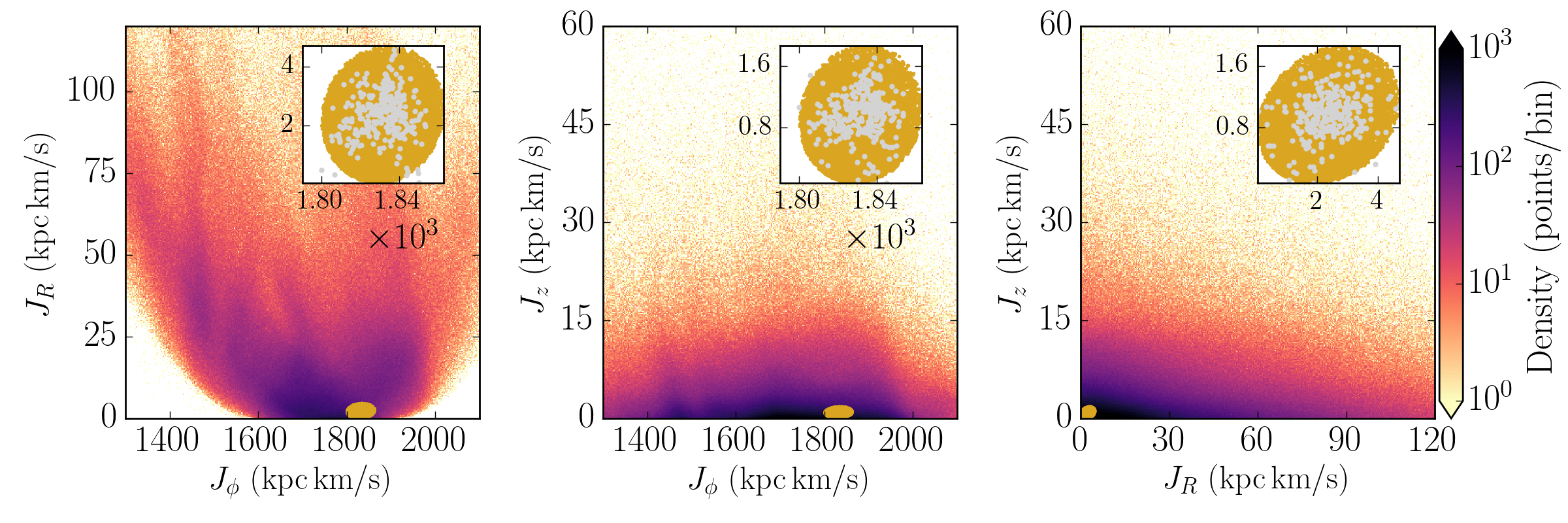}}
\put(-0.02,0){\includegraphics[width=\textwidth]{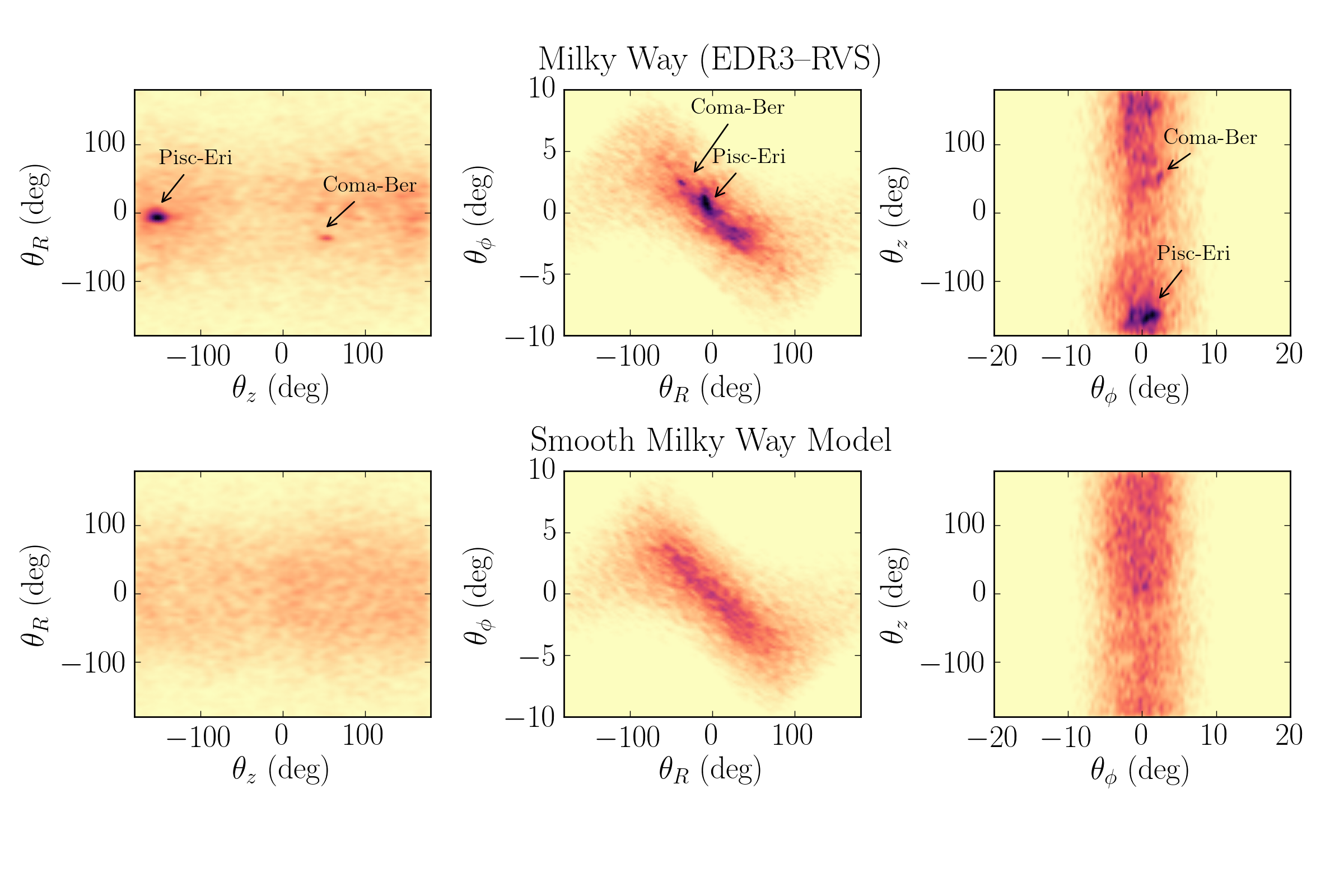}}
\end{picture}
\caption{Orbit patch and angle distribution for the Pisces Eridanus or Meingast 1 stream. The orbit patch contains $\sim$ 17\,000 stars whereas our FoF selection of the Pisces Eridanus stream has 300 stars. The group member's are shown with grey dots.
\newline
\textit{\textbf{Middle and lower panels}}: The angle distribution is less clumpy than the previous example cases we have discussed throughout this paper. However, at a first glance we can already identify two clusters: the most prominent one (Pisces Eridanus) located at ($\theta_{R}$, $\theta_{\phi}$, $\theta_{z}$) = (0, 0, -150) deg, and the second one is more diffuse but still clearly visible at ($\theta_{R}$, $\theta_{\phi}$, $\theta_{z}$) = (-30, 2, 50) deg, which is Coma Berenices. In the $\theta_{\phi}$--$\theta{z}$ angle projection we notice another pearl next to Pisces Eridanus, located at $\theta_{\phi} \sim -1$ deg. Stars in this pearl belong to the Collinder 135 and also UBC 7 cluster.\\
Distribution in angle space for the same orbit patch as shown above, but for a smooth-phase mixed distribution. This mock catalog shows the general features of the dataset, but with no clumping.}
\label{fig:A_ellipsoid_pisc_eri}
\end{center}
\end{figure*}

In the top panel of Fig.~\ref{fig:A_ellipsoid_pisc_eri} we show the corresponding orbit patch for this stream and also its angle distribution in the middle and lower panels. This stream appears accompanied by a second pearl, which we identify as Coma Berenices or Melotte 111 that is in the vicinity of the Sun with an estimated age of $\sim$ 700 Myr \citep{2019A&A...624L..11F, Tang2019}. The location of these pearls is clearly visible in the $\theta_{z}$--$\theta_{R}$ angle projection, with Pisces Eridanus located at $\theta_{\phi}$ (deg) $\sim 0$ and $\theta_{z}$ (deg) $\sim -150$. Coma Berenices is located at $\theta_{\phi}$\, (deg) $\sim 2$ and $\theta_{z}$\, (deg) $\sim 50$. On a closer inspection to the $\theta_{\phi}$--$\theta{z}$ angle projection we notice another pearl next to Pisces Eridanus. The two appear split in $\theta_{\phi}$ (deg) $\sim 0$, with the secondary pearl located at $\theta_{\phi}$ (deg)$\sim -1$. We find that stars in this third pearl belong to the cluster Collinder 135 and also UBC 7. The latter have been identified as a physical pair of clusters by \citet{2020A&A...642L...4K} with an age between 40--50 Myr. 

This case study illustrates an important point that will deserve follow-up: not all pearls on the same string will be approximately co-eval, as evidenced by the large age differences of the pearls here. This means that being pearls on the same string does not necessarily imply any form of 'common birth origin'.

\section{Offset orbit patch continued}

\begin{figure*}
\setlength{\unitlength}{\textwidth}
\begin{center}
\begin{picture}(1,1)
\put(0,0.65){\includegraphics[width=\textwidth]{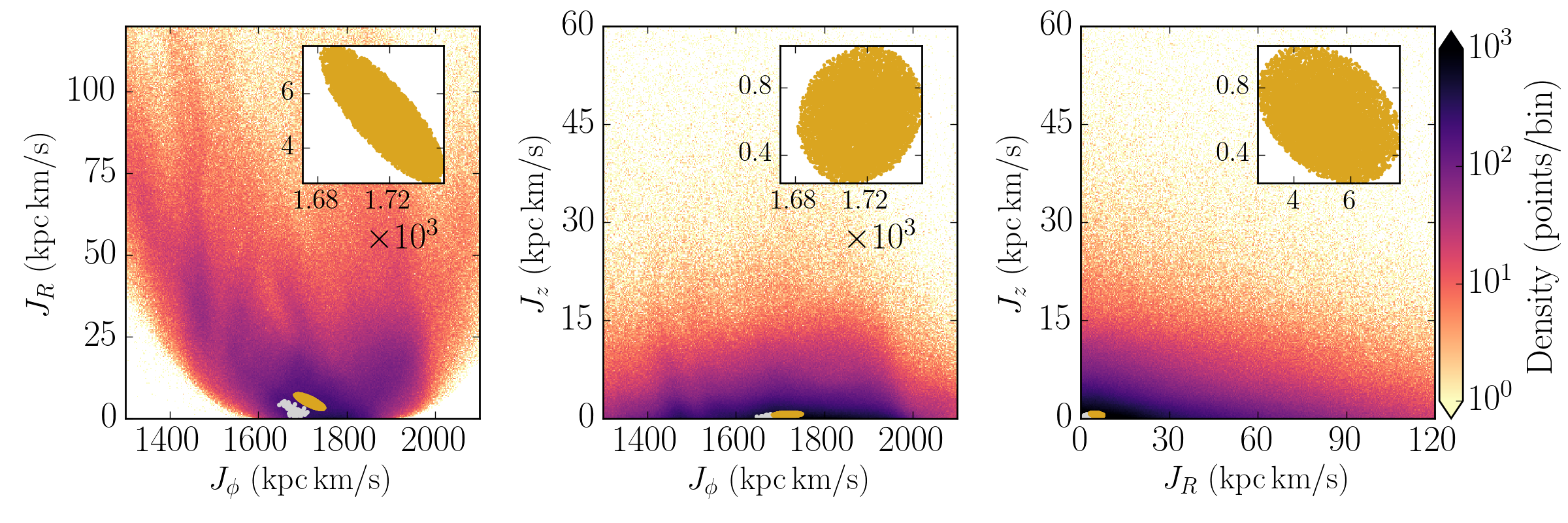}}
\put(-0.02,0){\includegraphics[width=\textwidth]{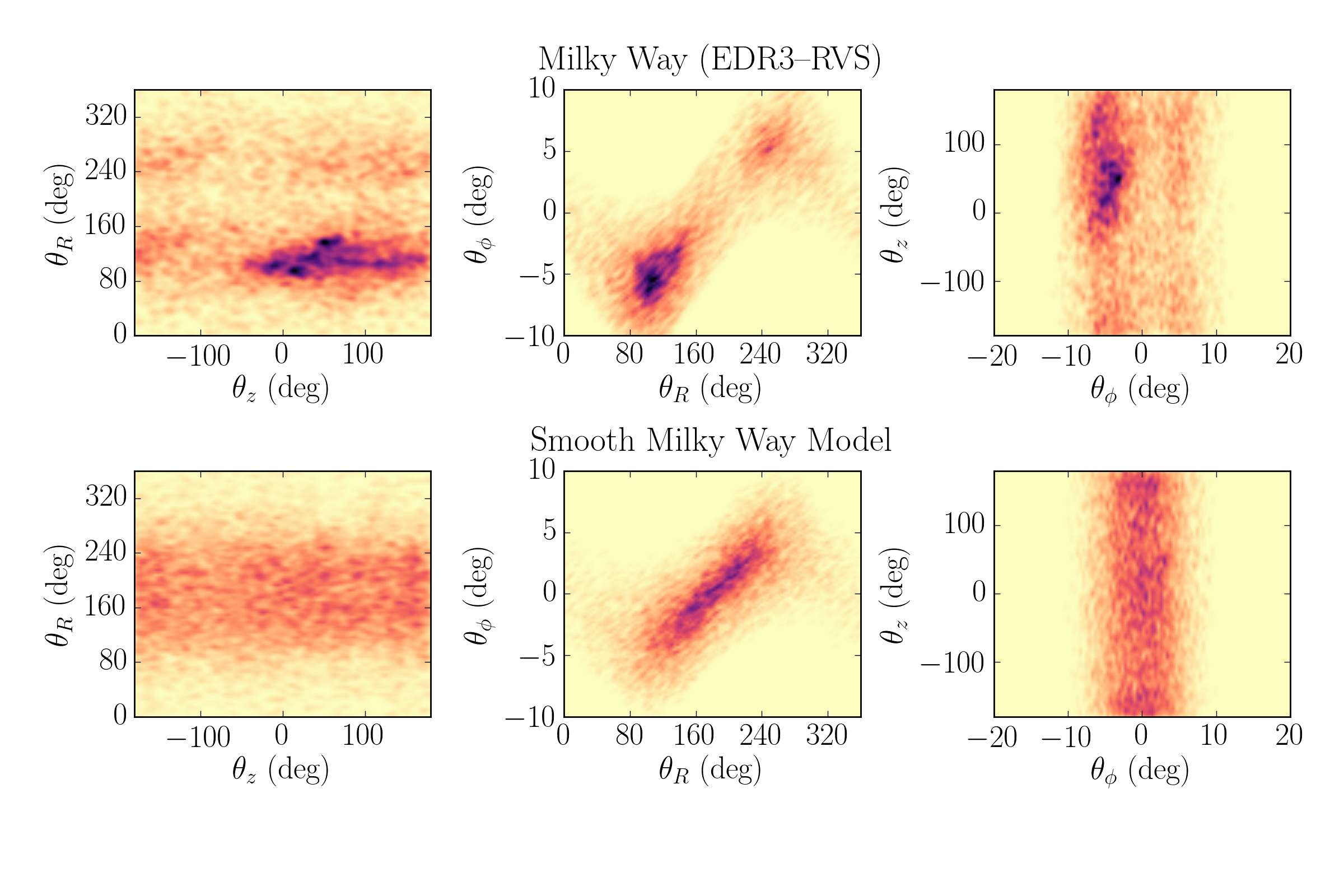}}
\end{picture}

\caption{Offset orbit patch selection. Here we show the \textit{FoF G19} orbit patch, but we shift its center $3\sigma$ (in $J_{R}, J_{z}, J_{\phi}$). Additionally we show the 2D projected (Gaussian) kernel density maps of the angle distribution ($\theta_{R}, \theta_{z}, \theta_{\phi}$) of the stars enclosed by the offset orbit patch. For the kernel width we selected the same size bandwidth for both the data and the mock catalog.\\
\textit{\textbf{Top panel}}: With yellow dots we show the new orbit patch location and in grey dots we show the original FoF group. The upper right inset in each panel shows a zoom-in into the orbit patch.\\
\textit{\textbf{Middle panel}}: We do see some overdensities and some sort of structure, however it does not resemble what we have shown in Figs.~\ref{fig:ellipsoid_example},~\ref{fig:angle_ellipsoid_example2} and~\ref{fig:angle_ellipsoid_example3}. We see that the distribution is not completely smooth. There is some large structure present approximately at $0 <\theta_{z}\, \rm{(deg)} < 100$ and $80<\theta_{R}\, \rm{(deg)} < 160$, which is expected as this offset orbit patch selection still falls in a high density region in action space. \\
\textit{\textbf{Lower panel}}: Distribution in angle space for the same orbit patch as shown above, but for a smooth-phase mixed distribution. This mock catalog shows the general features of the dataset, but with no clumping.}
\label{fig:G2_6actions_rn}
\end{center}
\end{figure*}

\begin{figure*}
\setlength{\unitlength}{\textwidth}
\begin{center}
\begin{picture}(1,1)
\put(0,0.65){\includegraphics[width=\textwidth]{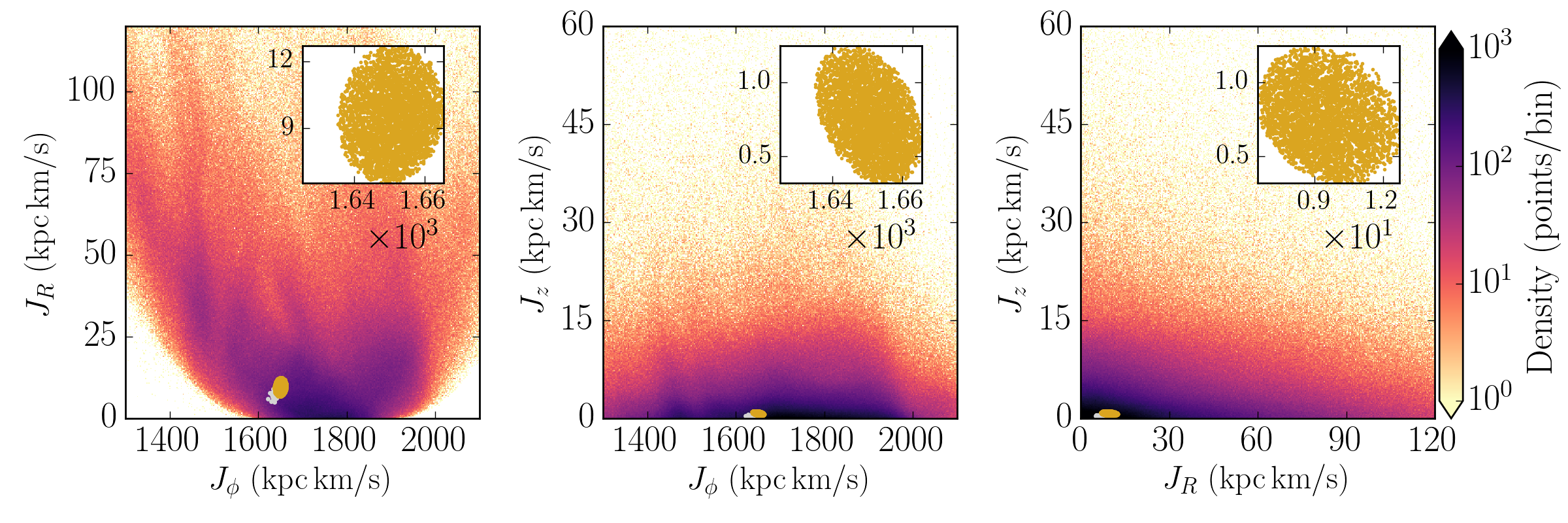}}
\put(-0.02,0){\includegraphics[width=\textwidth]{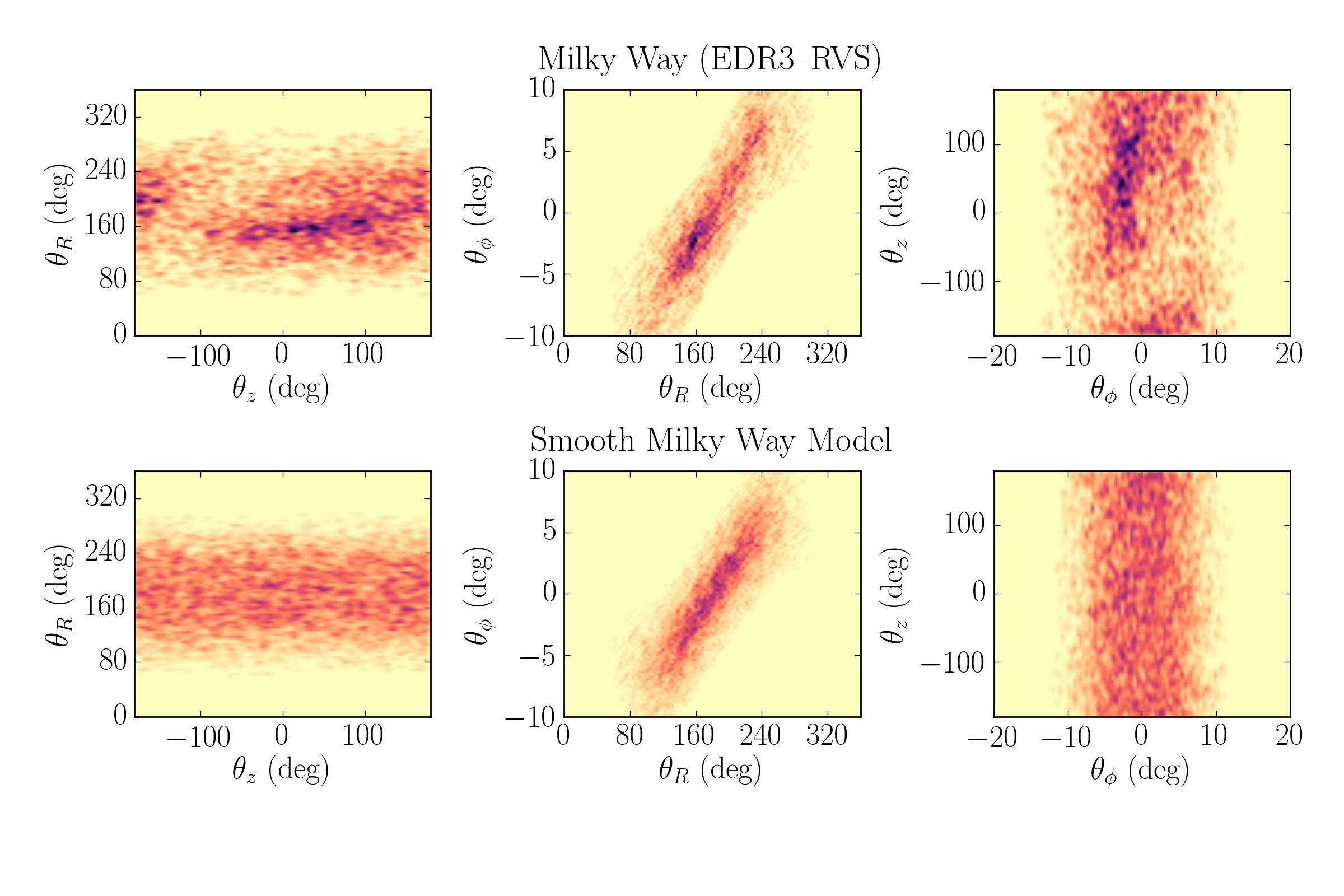}}
\end{picture}
\caption{Offset orbit patch selection. Here we show the \textit{FoF G40} orbit patch, but we shift its center $3\sigma$ (in $J_{R}, J_{z}, J_{\phi}$). Additionally we show the 2D projected kernel density maps of the angle distribution ($\theta_{R}, \theta_{z}, \theta_{\phi}$) of the stars enclosed by the offset orbit patch. For the kernel we selected the same size bandwidth for both the data and the mock catalog.
\newline
\textit{\textbf{Top panel}}: With yellow dots we show the new orbit patch location and in grey dots we show the original FoF group. The upper right inset in each panel shows a zoom-in into the orbit patch.
\newline
\textit{\textbf{Middle panel}}: We do see some overdensities and some sort of structure, however it does not resemble what we have shown in Figs.~\ref{fig:ellipsoid_example},~\ref{fig:angle_ellipsoid_example2} and~\ref{fig:angle_ellipsoid_example3}. We see that the distribution is not completely smooth. There is some large structure present approximately at $0 <\theta_{z}\, \rm{(deg)} < 100$ and $-10<\theta_{\phi}\, \rm{(deg)} < 0$, which is expected as this offset orbit patch selection still falls in a high density region in action space.
\newline
\textit{\textbf{Lower panel}}: Distribution in angle space for the same orbit patch as shown above, but for a smooth-phase mixed distribution. This mock catalog shows the general features of the dataset, but with no clumping.}
\label{fig:G3_2actions_rn}
\end{center}
\end{figure*}

Here we show two more examples of the offset orbit patch method already described in Section.~\ref{sec:random_ellipsoid}, for \textit{FoF G19} and \textit{FoF G40} in Figs.~\ref{fig:G2_6actions_rn} and ~\ref{fig:G3_2actions_rn}, respectively. Similar to the example for the \textit{FoF G12} orbit patch, we observe some orbital-phase overdensities present in both cases (mostly in Fig.~\ref{fig:G2_6actions_rn}), but nothing as structured as what we see in the original groups. Also we still expect to see some overdensities as we are selecting our offset orbit patch in very dense areas.

\section{Isolated groups}
\label{sec:isolated_groups}

\begin{figure*}
\begin{center}
\includegraphics[width=\textwidth]{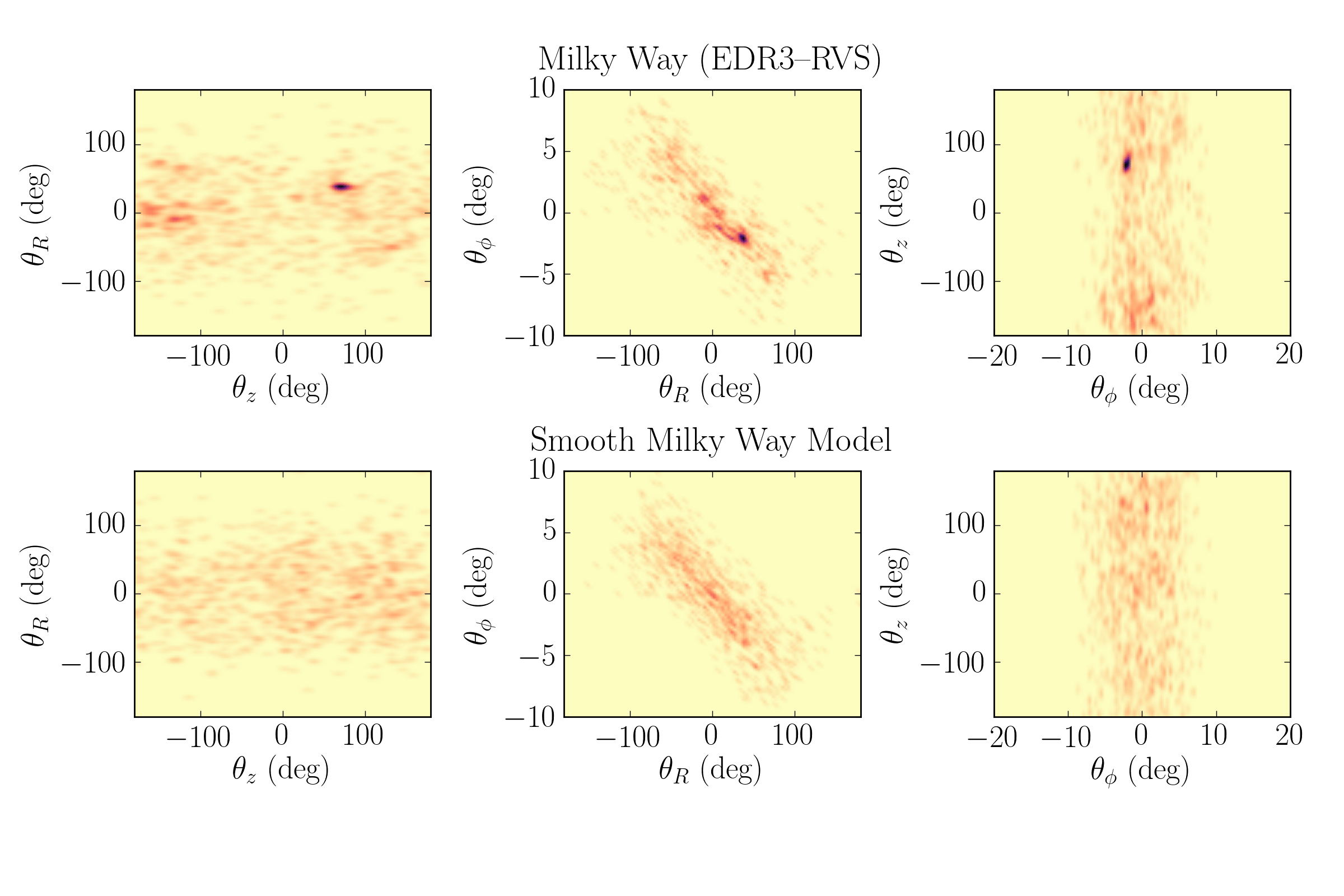}
\caption{2D projected kernel density maps of the angle distribution ($\theta_{R}, \theta_{z}, \theta_{\phi}$) of the stars enclosed by the orbit patch of one isolated cluster example.
This pearl located at ($\theta_{R}, \theta_{z}, \theta_{\phi})\,\rm(deg) \sim (60,80,-3)$  actually corresponds to a known cluster: Platais 3. The lower panel shows the smooth-phase mixed distribution that reproduces the main trends of the dataset where we don't see the cluster.}
\label{fig:G3_0angles}
\end{center}
\end{figure*}

In this work we have shown three example cases where the distribution in angle-space is highly structured, with many orbital phase overdensities or pearls. We observe that this is typical for many of the 55 groups where most of them have two or more cluster companions in the angle distribution. However we do have some cases that are isolated ($\sim$ 15\%). In Fig.~\ref{fig:G3_0angles} we show one example of the distribution in angle space for one of our isolated cases. And this isolated pearl is a known cluster: Platais 3. It has a presumed age of $\sim$ 830 Myr (log(age)=8.92, \citet{2018A&A...615A..12Y}), considerably older than many of the pearls discussed so far. A systematic determination of ages for the groups will be presented in a forthcoming paper. The fact that older pearls are more likely to be isolated could be the result of a low survival rate (dispersal) of most clusters and associations within a few 100 Myrs, but it could also show that groups considerably move away from their birth orbits within a few 100 Myrs, as expected from radial migration \citep{2018ApJ...865...96F}.

\section{Properties and Membership of the initial 55 groups}
In this section we present the properties and membership of the initial groups in Table~\ref{table:table1} and a subsample of the identified member stars of the 55 groups in Table~\ref{tab:vizier} (the machine-readable format is available in the online Journal).

\newpage
\begin{center}
\begin{longtable}{|l|l|l|l|l|l|l|l|}
\caption{Properties of the 55 initial groups.} \label{tab:long} \\

\hline \multicolumn{1}{|c|}{\textbf{Group ID}} & \multicolumn{1}{c|}{\textbf{$N_{members}$}} & \multicolumn{1}{c|}{\textbf{$\overline{J_{R}}$} (kpc km/s)} &
\multicolumn{1}{c|}{\textbf{$\overline{J_{z}}$} (kpc km/s)}&
\multicolumn{1}{c|}{\textbf{$\overline{J_{\phi}}$} (kpc km/s)}& \multicolumn{1}{c|}{\textbf{$\overline{\theta_{R}}$} (deg)} &
\multicolumn{1}{c|}{\textbf{$\overline{\theta_{z}}$} (deg)}& \multicolumn{1}{c|}{\textbf{$\overline{\theta_{\phi}}$} (deg)}\\ \hline 
\endfirsthead

\multicolumn{8}{c}%
{{\bfseries \tablename\ \thetable{} -- continued from previous page}} \\
\hline\multicolumn{1}{|c|}{\textbf{Group ID}} &
\multicolumn{1}{|c|}{\textbf{$N_{members}$}} & \multicolumn{1}{c|}{\textbf{$\overline{J_{R}}$} (kpc km/s)} & \multicolumn{1}{c|}{\textbf{$\overline{J_{z}}$} (kpc km/s)} &
\multicolumn{1}{c|}{\textbf{$\overline{J_{\phi}}$} (kpc km/s)}& \multicolumn{1}{c|}{\textbf{$\overline{\theta_{R}}$} (deg)} &
\multicolumn{1}{c|}{\textbf{$\overline{\theta_{z}}$} (deg)} &
\multicolumn{1}{c|}{\textbf{$\overline{\theta_{\phi}}$} (deg)}
\\
\hline 
\endhead

\hline \multicolumn{8}{|r|}{{Continued on next page}} \\ \hline
\endfoot
\hline \hline
\endlastfoot

1 & 83   &0.58  &0.61 & 1697.71 &-158.65 &17.09  & -2.87 \\
2 & 50   &0.82  &0.48 & 1693.89 &55.29   &96.38   & 2.1   \\
3 & 43   &0.87  &0.24 & 1814.44 &-17.63  &-163.69& -1.55 \\
4 & 70   &0.89  &0.19 & 1819.63 &-3.71   &-152.27 & -1.65 \\
5 & 48   &1.08  &0.63 & 1789.01 &-44.87  &57.7   & 2.11  \\
6 & 21   &1.27  &0.09 & 1736.38 &117.31  &57.45  & -3.34 \\
7 & 42   &1.47  &0.13 & 1723.99 &140.43  &146.58  & -4.74 \\
8 & 47   &1.82  &0.23 & 1755.3  &99.52   &164.6  & -4.47 \\
9 & 60   &1.84  &1.55 & 1796.04 &38.55   &-105.69 & -2.12 \\
10&  23  & 2.19 & 0.63 & 1795.45& 16.54  & -118.78&  -1.4 \\ 
11&  38  & 2.22 & 0.06 & 1697.96& 137.12 & 59.28 &  -0.11\\ 
12&  122 & 2.22 & 0.64 & 1671.73& -138.32& -10.57&  1.14 \\ 
13&  51  & 2.36 & 0.11 & 1804.01& 38.1   & 54.67 &  0.05 \\ 
14&  306 & 2.38 & 0.97 & 1831.74& -7.58  & -152.6&  0.91 \\ 
15&  23  & 2.42 & 0.64 & 1667.65& -163.39& 98.79 &  2.21 \\ 
16&  638 & 2.47 & 0.17 & 1677.98& -146.81& 69.97 &  1.53 \\ 
17&  39 & 2.54 & 0.25 & 1853.75& -7.8   & -133.9 &  0.31 \\ 
18&  69  & 2.68 & 0.77 & 1814.86& -37.65 & 53.75 &  2.51 \\ 
19&  231 & 2.72 & 0.23 & 1681.33& -149.47& 5.06  &  0.43 \\ 
20&  22  & 2.83 & 0.34 & 1792.87& 52.49  & 112.42&  -4.19\\ 
21&  40  & 3.05 & 1.2  & 1793.02& 33.97  & -88.5  &  -1.95\\ 
22&  45  & 3.11 & 0.18 & 1867.79& -26.78 & 92.58 &  2.4  \\ 
23&  26  & 3.13 & 0.78 & 1675.05& 171.99 & 61.99 &  1.55 \\ 
24&  28  & 3.23 & 0.12 & 1719.90 & 160.13&  16.99&  -1.78\\ 
25&  22 & 3.25 & 0.342& 1659.88& 130.46 & -34.06 &  -2.99\\ 
26&  41  & 3.33 & 0.21 & 1847.26& 37.92  & 71.0  &  -2.07\\ 
27&  90  & 3.49 & 0.14 & 1693.18& 165.17 & 70.88 &  -0.54\\ 
28&  24  & 3.61 & 0.13 & 1691.95& 166.78 & -89.21 &  -0.98\\ 
29&  91  & 3.62 & 0.07 & 1696.48& 162.15 & 30.3  &  -0.82\\ 
30&  86  & 3.71 & 1.76 & 1760.63& 122.23 & -149.5 &  -1.99\\ 
31&  52  & 3.87 & 0.41 & 1645.84& 120.84 & -61.01&  -0.76\\ 
32&  40  & 4.02 & 0.16 & 1807.60 & -12.23&  -169.8 & 0.67\\  
33&  32  & 4.26 & 0.18 & 1743.87& 137.28 & 124.42  &-5.76\\ 
34&  23  & 4.32 & 0.22 & 1608.60 & 140.45&  108.67 & -1.15 \\
35&  81  & 4.52 & 0.69 & 1709.20 & 136.01&  49.35  & -3.44 \\
36&  205  & 5.01 & 0.33 & 1651.03& 119.79 & -66.5   &-2.94  \\
37&  24  & 5.12 & 0.76 & 1626.60 & 166.35&  25.84  & -0.13 \\
38&  34  & 5.22 & 0.35 & 1714.87& 103.35 & -12.59  &-6.81  \\
39&  94  & 6.5  & 0.27 & 1635.65& 169.57 & 2.75    &-1.44  \\
40&  24  & 6.58 & 0.31 & 1635.33& 172.72 & 15.01   &-1.0   \\
41&  622 & 6.93 & 0.35 & 1651.24& -158.07& -114.25 &1.21   \\
42&  23  & 7.13 & 0.37 & 1719.75& 131.52 & 16.02    &-8.61  \\
43&  22  & 7.8  & 0.24 & 1746.80 & 80.48 &  -63.0   & -7.61 \\
44&  52  & 7.99 & 0.59 & 1738.90 & 103.75&  -0.54  & -6.97 \\
45&  39  & 10.34& 0.11 & 1914.97& 0.69   & -62.73   &-1.8   \\
46&  172 & 10.62& 0.37 & 1773.90 & 94.65 &  152.77 & -6.055\\
47&  52  & 13.54& 0.72 & 1757.58& 79.15  & -3.51   &-5.07  \\
48&  226 & 13.76& 0.58 & 1725.25& 99.45  & 84.57   &-9.3   \\
49&  260 & 14.1 & 0.24 & 1713.48& 96.62  & -15.29  &-9.13  \\
50&  31  & 14.82& 0.22 & 1831.94& -56.86 & -171.03  &10.55  \\
51&  48  & 16.29& 0.9  & 1566.01& 116.47 & -156.84 &-9.668 \\
52&  30  & 16.6 & 0.14 & 1882.98& -39.53 & 116.69  &6.84   \\
53&  84  & 17.79& 1.52 & 1910.88& -35.13 & 90.18   &7.31   \\
54&  20  & 18.36& 0.39 & 1924.46& -42.01 & 78.04    &6.46   \\
55&  118  & 40.95& 2.08 & 1649.94& -81.42 & -137.06  &16.55  \\

\end{longtable}
\label{table:table1}
\end{center}

\begin{deluxetable*}{cccccccc}
\tablenum{2}
\tablecaption{Subsample of the identified member stars of the 55 FoF groups. \label{tab:vizier}}
\tablewidth{0pt}
\tablehead{
\colhead{Group ID} & \colhead{Gaia EDR3 Source ID} & \colhead{$J_{R}$} & \colhead{$J_{z}$} & \colhead{$J_{\phi}$} & \colhead{$\theta_{R}$} & \colhead{$\theta_{z}$} & \colhead{$\theta_{\phi}$} \\
\colhead{Number} & \colhead{Number} & \colhead{(kpc km/s)} & \colhead{(kpc km/s)} & \colhead{(kpc km/s)} & \colhead{(deg)} & \colhead{(deg)} & \colhead{(deg)}
}
\decimalcolnumbers
\startdata
1 & 5339073605139436800 & 0.3411 & 0.6026 & 1702.3392 & -150.0429 & 17.4076 & -2.5263 \\
1 & 5338656477916975104 & 0.2858 & 0.5262 & 1706.7127 & -161.5242 & 17.9502 & -2.8299 \\
1 & 5338656310466146176 & 0.6125 & 0.6068 & 1694.0561 & -176.3901 & 15.9946 & -3.0948 \\
1 & 5338707472109552896 & 1.4188 & 0.5311 & 1675.8665 & -171.6802 & 17.8884 & -2.7933 \\
1 & 5338655722002740352 & 0.5046 & 0.4669 & 1698.2721 & -166.5107 & 18.8533 & -2.8405 \\
1 & 5338706544396496768 & 0.8148 & 0.4176 & 1688.9609 & -176.2026 & 19.2856 & -3.1229 \\
1 & 5338705719763247232 & 0.6474 & 0.7042 & 1693.7971 & -167.0601 & 15.3254 & -2.7961 \\
1 & 5338705509263621632 & 0.3022 & 0.6473 & 1705.4578 & -170.3867 & 15.4545 & -2.9893 \\
1 & 5338653454260022784 & 0.2951 & 0.6623 & 1704.5258 & 177.8306 & 14.4023 & -3.3224 \\
1 & 5338653282461326336 & 0.2057 & 0.5596 & 1709.5126 & -166.5654 & 16.6565 & -2.9947 \\
\enddata
\tablecomments{This table is available in machine-readable form.}
\end{deluxetable*}

\bibliography{pearls_on_a_string}{}

\begin{thebibliography}{}
\expandafter\ifx\csname natexlab\endcsname\relax\def\natexlab#1{#1}\fi
\providecommand{\url}[1]{\href{#1}{#1}}
\providecommand{\dodoi}[1]{doi:~\href{http://doi.org/#1}{\nolinkurl{#1}}}
\providecommand{\doeprint}[1]{\href{http://ascl.net/#1}{\nolinkurl{http://ascl.net/#1}}}
\providecommand{\doarXiv}[1]{\href{https://arxiv.org/abs/#1}{\nolinkurl{https://arxiv.org/abs/#1}}}

\bibitem[{{Alves} {et~al.}(2020){Alves}, {Zucker}, {Goodman}, {Speagle},
  {Meingast}, {Robitaille}, {Finkbeiner}, {Schlafly}, \& {Green}}]{Alves2020}
{Alves}, J., {Zucker}, C., {Goodman}, A.~A., {et~al.} 2020, \nat, 578, 237,
  \dodoi{10.1038/s41586-019-1874-z}

\bibitem[{{Bailer-Jones} {et~al.}(2020){Bailer-Jones}, {Rybizki}, {Fouesneau},
  {Demleitner}, \& {Andrae}}]{2020arXiv201205220B}
{Bailer-Jones}, C.~A.~L., {Rybizki}, J., {Fouesneau}, M., {Demleitner}, M., \&
  {Andrae}, R. 2020, arXiv e-prints, arXiv:2012.05220.
\newblock \doarXiv{2012.05220}

\bibitem[{{Bastian}(2019)}]{2019A&A...630L...8B}
{Bastian}, U. 2019, \aap, 630, L8, \dodoi{10.1051/0004-6361/201936595}

\bibitem[{{Binney}(2012)}]{2012MNRAS.426.1324B}
{Binney}, J. 2012, \mnras, 426, 1324, \dodoi{10.1111/j.1365-2966.2012.21757.x}

\bibitem[{{Binney} \& {Tremaine}(2008)}]{binney2008}
{Binney}, J., \& {Tremaine}, S. 2008, {Galactic Dynamics: Second Edition}
  (Princeton University Press)

\bibitem[{{Boesgaard} {et~al.}(2013){Boesgaard}, {Roper}, \&
  {Lum}}]{2013ApJ...775...58B}
{Boesgaard}, A.~M., {Roper}, B.~W., \& {Lum}, M.~G. 2013, \apj, 775, 58,
  \dodoi{10.1088/0004-637X/775/1/58}

\bibitem[{{Bonatto} \& {Bica}(2010)}]{2010MNRAS.403..996B}
{Bonatto}, C., \& {Bica}, E. 2010, \mnras, 403, 996,
  \dodoi{10.1111/j.1365-2966.2009.16177.x}

\bibitem[{{Bovy}(2015)}]{galpy}
{Bovy}, J. 2015, \apjs, 216, 29, \dodoi{10.1088/0067-0049/216/2/29}

\bibitem[{{Brandt} \& {Huang}(2015)}]{2015ApJ...807...24B}
{Brandt}, T.~D., \& {Huang}, C.~X. 2015, \apj, 807, 24,
  \dodoi{10.1088/0004-637X/807/1/24}

\bibitem[{{Bressan} {et~al.}(2012){Bressan}, {Marigo}, {Girardi}, {Salasnich},
  {Dal Cero}, {Rubele}, \& {Nanni}}]{2012MNRAS.427..127B}
{Bressan}, A., {Marigo}, P., {Girardi}, L., {et~al.} 2012, \mnras, 427, 127,
  \dodoi{10.1111/j.1365-2966.2012.21948.x}

\bibitem[{{Cantat-Gaudin} \& {Anders}(2020)}]{2020A&A...633A..99C}
{Cantat-Gaudin}, T., \& {Anders}, F. 2020, \aap, 633, A99,
  \dodoi{10.1051/0004-6361/201936691}

\bibitem[{{Carraro} \& {Patat}(2001)}]{2001A&A...379..136C}
{Carraro}, G., \& {Patat}, F. 2001, \aap, 379, 136,
  \dodoi{10.1051/0004-6361:20011314}

\bibitem[{{Conrad} {et~al.}(2017){Conrad}, {Scholz}, {Kharchenko}, {Piskunov},
  {R{\"o}ser}, {Schilbach}, {de Jong}, {Schnurr}, {Steinmetz}, {Grebel},
  {Zwitter}, {Bienaym{\'e}}, {Bland-Hawthorn}, {Gibson}, {Gilmore},
  {Kordopatis}, {Kunder}, {Navarro}, {Parker}, {Reid}, {Seabroke}, {Siviero},
  {Watson}, \& {Wyse}}]{2017A&A...600A.106C}
{Conrad}, C., {Scholz}, R.~D., {Kharchenko}, N.~V., {et~al.} 2017, \aap, 600,
  A106, \dodoi{10.1051/0004-6361/201630012}

\bibitem[{{Coronado} {et~al.}(2018){Coronado}, {Rix}, \&
  {Trick}}]{2018MNRAS.481.2970C}
{Coronado}, J., {Rix}, H.-W., \& {Trick}, W.~H. 2018, \mnras, 481, 2970,
  \dodoi{10.1093/mnras/sty2468}

\bibitem[{{Coronado} {et~al.}(2020){Coronado}, {Rix}, {Trick}, {El-Badry},
  {Rybizki}, \& {Xiang}}]{2020MNRAS.495.4098C}
{Coronado}, J., {Rix}, H.-W., {Trick}, W.~H., {et~al.} 2020, \mnras, 495, 4098,
  \dodoi{10.1093/mnras/staa1358}

\bibitem[{{Curtis} {et~al.}(2019){Curtis}, {Ag{\"u}eros}, {Mamajek}, {Wright},
  \& {Cummings}}]{2019AJ....158...77C}
{Curtis}, J.~L., {Ag{\"u}eros}, M.~A., {Mamajek}, E.~E., {Wright}, J.~T., \&
  {Cummings}, J.~D. 2019, \aj, 158, 77, \dodoi{10.3847/1538-3881/ab2899}

\bibitem[{{de La Fuente Marcos} \& {de La Fuente
  Marcos}(2009)}]{2009A&A...500L..13D}
{de La Fuente Marcos}, R., \& {de La Fuente Marcos}, C. 2009, \aap, 500, L13,
  \dodoi{10.1051/0004-6361/200912297}

\bibitem[{{Dias} {et~al.}(2002){Dias}, {Alessi}, {Moitinho}, \&
  {L{\'e}pine}}]{2002A&A...389..871D}
{Dias}, W.~S., {Alessi}, B.~S., {Moitinho}, A., \& {L{\'e}pine}, J.~R.~D. 2002,
  \aap, 389, 871, \dodoi{10.1051/0004-6361:20020668}

\bibitem[{{Dieball} {et~al.}(2002){Dieball}, {M{\"u}ller}, \&
  {Grebel}}]{2002A&A...391..547D}
{Dieball}, A., {M{\"u}ller}, H., \& {Grebel}, E.~K. 2002, \aap, 391, 547,
  \dodoi{10.1051/0004-6361:20020815}

\bibitem[{{Douglas} {et~al.}(2014){Douglas}, {Ag{\"u}eros}, {Covey}, {Bowsher},
  {Bochanski}, {Cargile}, {Kraus}, {Law}, {Lemonias}, {Arce}, {Fierroz}, \&
  {Kundert}}]{2014ApJ...795..161D}
{Douglas}, S.~T., {Ag{\"u}eros}, M.~A., {Covey}, K.~R., {et~al.} 2014, \apj,
  795, 161, \dodoi{10.1088/0004-637X/795/2/161}

\bibitem[{{Eggen}(1959)}]{1959Obs....79..143E}
{Eggen}, O.~J. 1959, The Observatory, 79, 143

\bibitem[{{Eggen}(1960)}]{1960MNRAS.120..540E}
---. 1960, \mnras, 120, 540, \dodoi{10.1093/mnras/120.6.540}

\bibitem[{{Eilers} {et~al.}(2019){Eilers}, {Hogg}, {Rix}, \&
  {Ness}}]{2019ApJ...871..120E}
{Eilers}, A.-C., {Hogg}, D.~W., {Rix}, H.-W., \& {Ness}, M.~K. 2019, \apj, 871,
  120, \dodoi{10.3847/1538-4357/aaf648}

\bibitem[{{Elmegreen} \& {Efremov}(1996)}]{1996ApJ...466..802E}
{Elmegreen}, B.~G., \& {Efremov}, Y.~N. 1996, \apj, 466, 802,
  \dodoi{10.1086/177554}

\bibitem[{{Elmegreen} \& {Falgarone}(1996)}]{1996ApJ...471..816E}
{Elmegreen}, B.~G., \& {Falgarone}, E. 1996, \apj, 471, 816,
  \dodoi{10.1086/178009}

\bibitem[{{Fabricius} {et~al.}(2020){Fabricius}, {Luri}, {Arenou}, {Babusiaux},
  {Helmi}, {Muraveva}, {Reyl{\'e}}, {Spoto}, {Vallenari}, {Antoja}, {Balbinot},
  {Barache}, {Bauchet}, {Bragaglia}, {Busonero}, {Cantat-Gaudin}, {Carrasco},
  {Diakit{\'e}}, {Fabrizio}, {Figueras}, {Garcia-Gutierrez}, {Garofalo},
  {Jordi}, {Kervella}, {Khanna}, {Leclerc}, {Licata}, {Lambert}, {Marrese},
  {Masip}, {Ramos}, {Robichon}, {Robin}, {Romero-G{\'o}mez}, {Rubele}, \&
  {Weiler}}]{2020arXiv201206242F}
{Fabricius}, C., {Luri}, X., {Arenou}, F., {et~al.} 2020, arXiv e-prints,
  arXiv:2012.06242.
\newblock \doarXiv{2012.06242}

\bibitem[{{Frankel} {et~al.}(2018){Frankel}, {Rix}, {Ting}, {Ness}, \&
  {Hogg}}]{2018ApJ...865...96F}
{Frankel}, N., {Rix}, H.-W., {Ting}, Y.-S., {Ness}, M., \& {Hogg}, D.~W. 2018,
  \apj, 865, 96, \dodoi{10.3847/1538-4357/aadba5}

\bibitem[{{Fritzewski} {et~al.}(2019){Fritzewski}, {Barnes}, {James}, {Geller},
  {Meibom}, \& {Strassmeier}}]{2019A&A...622A.110F}
{Fritzewski}, D.~J., {Barnes}, S.~A., {James}, D.~J., {et~al.} 2019, \aap, 622,
  A110, \dodoi{10.1051/0004-6361/201833587}

\bibitem[{{Fujimoto} \& {Kumai}(1997)}]{1997AJ....113..249F}
{Fujimoto}, M., \& {Kumai}, Y. 1997, \aj, 113, 249, \dodoi{10.1086/118249}

\bibitem[{{F{\"u}rnkranz} {et~al.}(2019){F{\"u}rnkranz}, {Meingast}, \&
  {Alves}}]{2019A&A...624L..11F}
{F{\"u}rnkranz}, V., {Meingast}, S., \& {Alves}, J. 2019, \aap, 624, L11,
  \dodoi{10.1051/0004-6361/201935293}

\bibitem[{{Gagn{\'e}} \& {Faherty}(2018)}]{2018ApJ...862..138G}
{Gagn{\'e}}, J., \& {Faherty}, J.~K. 2018, \apj, 862, 138,
  \dodoi{10.3847/1538-4357/aaca2e}

\bibitem[{{Gagn{\'e}} {et~al.}(2018){Gagn{\'e}}, {Faherty}, \&
  {Mamajek}}]{2018ApJ...865..136G}
{Gagn{\'e}}, J., {Faherty}, J.~K., \& {Mamajek}, E.~E. 2018, \apj, 865, 136,
  \dodoi{10.3847/1538-4357/aadaed}

\bibitem[{{Gossage} {et~al.}(2018){Gossage}, {Conroy}, {Dotter}, {Choi},
  {Rosenfield}, {Cargile}, \& {Dolphin}}]{2018ApJ...863...67G}
{Gossage}, S., {Conroy}, C., {Dotter}, A., {et~al.} 2018, \apj, 863, 67,
  \dodoi{10.3847/1538-4357/aad0a0}

\bibitem[{{Gusev} \& {Efremov}(2013)}]{2013MNRAS.434..313G}
{Gusev}, A.~S., \& {Efremov}, Y.~N. 2013, \mnras, 434, 313,
  \dodoi{10.1093/mnras/stt1019}

\bibitem[{{Hawkins} {et~al.}(2020){Hawkins}, {Lucey}, \&
  {Curtis}}]{2020MNRAS.496.2422H}
{Hawkins}, K., {Lucey}, M., \& {Curtis}, J. 2020, \mnras, 496, 2422,
  \dodoi{10.1093/mnras/staa1673}

\bibitem[{{Katz} {et~al.}(2019){Katz}, {Sartoretti}, {Cropper}, {Panuzzo},
  {Seabroke}, {Viala}, {Benson}, {Blomme}, {Jasniewicz}, {Jean-Antoine},
  {Huckle}, {Smith}, {Baker}, {Crifo}, {Damerdji}, {David}, {Dolding},
  {Fr{\'e}mat}, {Gosset}, {Guerrier}, {Guy}, {Haigron}, {Jan{\ss}en},
  {Marchal}, {Plum}, {Soubiran}, {Th{\'e}venin}, {Ajaj}, {Allende Prieto},
  {Babusiaux}, {Boudreault}, {Chemin}, {Delle Luche}, {Fabre}, {Gueguen},
  {Hambly}, {Lasne}, {Meynadier}, {Pailler}, {Panem}, {Royer}, {Tauran},
  {Zurbach}, {Zwitter}, {Arenou}, {Bossini}, {Gerssen}, {G{\'o}mez},
  {Lemaitre}, {Leclerc}, {Morel}, {Munari}, {Turon}, {Vallenari}, \&
  {{\v{Z}}erjal}}]{2019A&A...622A.205K}
{Katz}, D., {Sartoretti}, P., {Cropper}, M., {et~al.} 2019, \aap, 622, A205,
  \dodoi{10.1051/0004-6361/201833273}

\bibitem[{{Kharchenko} {et~al.}(2005){Kharchenko}, {Piskunov}, {R{\"o}ser},
  {Schilbach}, \& {Scholz}}]{2005A&A...438.1163K}
{Kharchenko}, N.~V., {Piskunov}, A.~E., {R{\"o}ser}, S., {Schilbach}, E., \&
  {Scholz}, R.~D. 2005, \aap, 438, 1163, \dodoi{10.1051/0004-6361:20042523}

\bibitem[{{Kontizas} {et~al.}(1993){Kontizas}, {Kontizas}, \&
  {Michalitsianos}}]{1993A&A...267...59K}
{Kontizas}, E., {Kontizas}, M., \& {Michalitsianos}, A. 1993, \aap, 267, 59

\bibitem[{{Kounkel} \& {Covey}(2019)}]{2019AJ....158..122K}
{Kounkel}, M., \& {Covey}, K. 2019, \aj, 158, 122,
  \dodoi{10.3847/1538-3881/ab339a}

\bibitem[{{Kov{\'a}cs} {et~al.}(2014){Kov{\'a}cs}, {Hartman}, {Bakos}, {Quinn},
  {Penev}, {Latham}, {Bhatti}, {Csubry}, \& {de
  Val-Borro}}]{2014MNRAS.442.2081K}
{Kov{\'a}cs}, G., {Hartman}, J.~D., {Bakos}, G.~{\'A}., {et~al.} 2014, \mnras,
  442, 2081, \dodoi{10.1093/mnras/stu946}

\bibitem[{{Kovaleva} {et~al.}(2020){Kovaleva}, {Ishchenko}, {Postnikova},
  {Berczik}, {Piskunov}, {Kharchenko}, {Polyachenko}, {Reffert}, {Sysoliatina},
  \& {Just}}]{2020A&A...642L...4K}
{Kovaleva}, D.~A., {Ishchenko}, M., {Postnikova}, E., {et~al.} 2020, \aap, 642,
  L4, \dodoi{10.1051/0004-6361/202039215}

\bibitem[{{Krumholz} {et~al.}(2019){Krumholz}, {McKee}, \&
  {Bland-Hawthorn}}]{2019ARA&A..57..227K}
{Krumholz}, M.~R., {McKee}, C.~F., \& {Bland-Hawthorn}, J. 2019, \araa, 57,
  227, \dodoi{10.1146/annurev-astro-091918-104430}

\bibitem[{{Lindegren} {et~al.}(2020){Lindegren}, {Klioner}, {Hern{\'a}ndez},
  {Bombrun}, {Ramos-Lerate}, {Steidelm{\"u}ller}, {Bastian}, {Biermann}, {de
  Torres}, {Gerlach}, {Geyer}, {Hilger}, {Hobbs}, {Lammers}, {McMillan},
  {Stephenson}, {Casta{\~n}eda}, {Davidson}, {Fabricius}, {Gracia-Abril},
  {Portell}, {Rowell}, {Teyssier}, {Torra}, {Bartolom{\'e}}, {Clotet},
  {Garralda}, {Gonz{\'a}lez-Vidal}, {Torra}, {Abbas}, {Altmann}, {Anglada
  Varela}, {Balaguer-N{\'u}{\~n}ez}, {Balog}, {Barache}, {Becciani}, {Bernet},
  {Bertone}, {Bianchi}, {Bouquillon}, {Brown}, {Bucciarelli}, {Busonero},
  {Butkevich}, {Buzzi}, {Cancelliere}, {Carlucci}, {Charlot}, {Cioni},
  {Crosta}, {Crowley}, {del Peloso}, {del Pozo}, {Drimmel}, {Esquej}, {Fienga},
  {Fraile}, {Gai}, {Garcia-Reinaldos}, {Guerra}, {Hambly}, {Hauser},
  {Jan{\ss}en}, {Jordan}, {Kostrzewa-Rutkowska}, {Lattanzi}, {Liao}, {Licata},
  {Lister}, {L{\"o}ffler}, {Marchant}, {Masip}, {Mignard}, {Mints}, {Molina},
  {Mora}, {Morbidelli}, {Murphy}, {Pagani}, {Panuzzo}, {Pe{\~n}alosa Esteller},
  {Poggio}, {Re Fiorentin}, {Riva}, {Sagrist{\`a} Sell{\'e}s}, {Sanchez
  Gimenez}, {Sarasso}, {Sciacca}, {Siddiqui}, {Smart}, {Souami}, {Spagna},
  {Steele}, {Taris}, {Utrilla}, {van Reeven}, \&
  {Vecchiato}}]{2020arXiv201203380L}
{Lindegren}, L., {Klioner}, S.~A., {Hern{\'a}ndez}, J., {et~al.} 2020, arXiv
  e-prints, arXiv:2012.03380.
\newblock \doarXiv{2012.03380}

\bibitem[{{Meingast} {et~al.}(2019){Meingast}, {Alves}, \&
  {F{\"u}rnkranz}}]{2019A&A...622L..13M}
{Meingast}, S., {Alves}, J., \& {F{\"u}rnkranz}, V. 2019, \aap, 622, L13,
  \dodoi{10.1051/0004-6361/201834950}

\bibitem[{{Meingast} {et~al.}(2021){Meingast}, {Alves}, \&
  {Rottensteiner}}]{2021A&A...645A..84M}
{Meingast}, S., {Alves}, J., \& {Rottensteiner}, A. 2021, \aap, 645, A84,
  \dodoi{10.1051/0004-6361/202038610}

\bibitem[{{Mucciarelli} {et~al.}(2012){Mucciarelli}, {Origlia}, {Ferraro},
  {Bellazzini}, \& {Lanzoni}}]{2012ApJ...746L..19M}
{Mucciarelli}, A., {Origlia}, L., {Ferraro}, F.~R., {Bellazzini}, M., \&
  {Lanzoni}, B. 2012, \apjl, 746, L19, \dodoi{10.1088/2041-8205/746/2/L19}

\bibitem[{{Palma} {et~al.}(2016){Palma}, {Gramajo}, {Clari{\'a}}, {Lares},
  {Geisler}, \& {Ahumada}}]{2016A&A...586A..41P}
{Palma}, T., {Gramajo}, L.~V., {Clari{\'a}}, J.~J., {et~al.} 2016, \aap, 586,
  A41, \dodoi{10.1051/0004-6361/201527305}

\bibitem[{{Perryman} {et~al.}(1998){Perryman}, {Brown}, {Lebreton}, {Gomez},
  {Turon}, {Cayrel de Strobel}, {Mermilliod}, {Robichon}, {Kovalevsky}, \&
  {Crifo}}]{1998A&A...331...81P}
{Perryman}, M.~A.~C., {Brown}, A.~G.~A., {Lebreton}, Y., {et~al.} 1998, \aap,
  331, 81.
\newblock \doarXiv{astro-ph/9707253}

\bibitem[{{Pietrzynski} \& {Udalski}(2000)}]{2000AcA....50..355P}
{Pietrzynski}, G., \& {Udalski}, A. 2000, \actaa, 50, 355.
\newblock \doarXiv{astro-ph/0010294}

\bibitem[{{Ratzenb{\"o}ck} {et~al.}(2020){Ratzenb{\"o}ck}, {Meingast}, {Alves},
  {M{\"o}ller}, \& {Bomze}}]{2020A&A...639A..64R}
{Ratzenb{\"o}ck}, S., {Meingast}, S., {Alves}, J., {M{\"o}ller}, T., \&
  {Bomze}, I. 2020, \aap, 639, A64, \dodoi{10.1051/0004-6361/202037591}

\bibitem[{{Reid} {et~al.}(2014){Reid}, {Menten}, {Brunthaler}, {Zheng}, {Dame},
  {Xu}, {Wu}, {Zhang}, {Sanna}, {Sato}, {Hachisuka}, {Choi}, {Immer},
  {Moscadelli}, {Rygl}, \& {Bartkiewicz}}]{2014ApJ...783..130R}
{Reid}, M.~J., {Menten}, K.~M., {Brunthaler}, A., {et~al.} 2014, \apj, 783,
  130, \dodoi{10.1088/0004-637X/783/2/130}

\bibitem[{{Rybizki} {et~al.}(2018){Rybizki}, {Demleitner}, {Fouesneau},
  {Bailer-Jones}, {Rix}, \& {Andrae}}]{2018PASP..130g4101R}
{Rybizki}, J., {Demleitner}, M., {Fouesneau}, M., {et~al.} 2018, \pasp, 130,
  074101, \dodoi{10.1088/1538-3873/aabd70}

\bibitem[{{Rybizki} {et~al.}(2021){Rybizki}, {Rix}, {Demleitner},
  {Bailer-Jones}, \& {Cooper}}]{Rybizki2021}
{Rybizki}, J., {Rix}, H.-W., {Demleitner}, M., {Bailer-Jones}, C. A.~L., \&
  {Cooper}, W.~J. 2021, \mnras, 500, 397, \dodoi{10.1093/mnras/staa3089}

\bibitem[{{Rybizki} {et~al.}(2020){Rybizki}, {Demleitner}, {Bailer-Jones},
  {Tio}, {Cantat-Gaudin}, {Fouesneau}, {Chen}, {Andrae}, {Girardi}, \&
  {Sharma}}]{2020PASP..132g4501R}
{Rybizki}, J., {Demleitner}, M., {Bailer-Jones}, C., {et~al.} 2020, \pasp, 132,
  074501, \dodoi{10.1088/1538-3873/ab8cb0}

\bibitem[{{Salaris} {et~al.}(2004){Salaris}, {Weiss}, \&
  {Percival}}]{2004A&A...414..163S}
{Salaris}, M., {Weiss}, A., \& {Percival}, S.~M. 2004, \aap, 414, 163,
  \dodoi{10.1051/0004-6361:20031578}

\bibitem[{{Sanders} \& {Binney}(2015)}]{2015MNRAS.449.3479S}
{Sanders}, J.~L., \& {Binney}, J. 2015, \mnras, 449, 3479,
  \dodoi{10.1093/mnras/stv578}

\bibitem[{{Sch{\"o}nrich} {et~al.}(2010){Sch{\"o}nrich}, {Binney}, \&
  {Dehnen}}]{2010MNRAS.403.1829S}
{Sch{\"o}nrich}, R., {Binney}, J., \& {Dehnen}, W. 2010, \mnras, 403, 1829,
  \dodoi{10.1111/j.1365-2966.2010.16253.x}

\bibitem[{{Sim} {et~al.}(2019){Sim}, {Lee}, {Ann}, \&
  {Kim}}]{2019JKAS...52..145S}
{Sim}, G., {Lee}, S.~H., {Ann}, H.~B., \& {Kim}, S. 2019, Journal of Korean
  Astronomical Society, 52, 145, \dodoi{10.5303/JKAS.2019.52.5.145}

\bibitem[{{Soubiran} {et~al.}(2018){Soubiran}, {Cantat-Gaudin},
  {Romero-G{\'o}mez}, {Casamiquela}, {Jordi}, {Vallenari}, {Antoja},
  {Balaguer-N{\'u}{\~n}ez}, {Bossini}, {Bragaglia}, {Carrera}, {Castro-Ginard},
  {Figueras}, {Heiter}, {Katz}, {Krone-Martins}, {Le Campion}, {Moitinho}, \&
  {Sordo}}]{2018A&A...619A.155S}
{Soubiran}, C., {Cantat-Gaudin}, T., {Romero-G{\'o}mez}, M., {et~al.} 2018,
  \aap, 619, A155, \dodoi{10.1051/0004-6361/201834020}

\bibitem[{{Spina} {et~al.}(2017){Spina}, {Randich}, {Magrini}, {Jeffries},
  {Friel}, {Sacco}, {Pancino}, {Bonito}, {Bravi}, {Franciosini}, {Klutsch},
  {Montes}, {Gilmore}, {Vallenari}, {Bensby}, {Bragaglia}, {Flaccomio},
  {Koposov}, {Korn}, {Lanzafame}, {Smiljanic}, {Bayo}, {Carraro}, {Casey},
  {Costado}, {Damiani}, {Donati}, {Frasca}, {Hourihane}, {Jofr{\'e}}, {Lewis},
  {Lind}, {Monaco}, {Morbidelli}, {Prisinzano}, {Sousa}, {Worley}, \&
  {Zaggia}}]{2017A&A...601A..70S}
{Spina}, L., {Randich}, S., {Magrini}, L., {et~al.} 2017, \aap, 601, A70,
  \dodoi{10.1051/0004-6361/201630078}

\bibitem[{{Tang} {et~al.}(2019){Tang}, {Pang}, {Yuan}, {Chen}, {Hong},
  {Goldman}, {Just}, {Shukirgaliyev}, \& {Lin}}]{Tang2019}
{Tang}, S.-Y., {Pang}, X., {Yuan}, Z., {et~al.} 2019, \apj, 877, 12,
  \dodoi{10.3847/1538-4357/ab13b0}

\bibitem[{{Trick} {et~al.}(2017){Trick}, {Bovy}, {D'Onghia}, \&
  {Rix}}]{Trick_2017}
{Trick}, W.~H., {Bovy}, J., {D'Onghia}, E., \& {Rix}, H.-W. 2017, \apj, 839,
  61, \dodoi{10.3847/1538-4357/aa67db}

\bibitem[{{Trick} {et~al.}(2019){Trick}, {Coronado}, \&
  {Rix}}]{2019MNRAS.484.3291T}
{Trick}, W.~H., {Coronado}, J., \& {Rix}, H.-W. 2019, \mnras, 484, 3291,
  \dodoi{10.1093/mnras/stz209}

\bibitem[{{Yen} {et~al.}(2018){Yen}, {Reffert}, {Schilbach}, {R{\"o}ser},
  {Kharchenko}, \& {Piskunov}}]{2018A&A...615A..12Y}
{Yen}, S.~X., {Reffert}, S., {Schilbach}, E., {et~al.} 2018, \aap, 615, A12,
  \dodoi{10.1051/0004-6361/201731905}

\end{thebibliography}
\bibliographystyle{aasjournal}



\end{document}